\documentclass[12pt,a4paper]{article}
\pdfoutput=1
\usepackage{jheppub}  
\usepackage{amssymb} 
\usepackage{amsmath}
\usepackage{mathtools}
\usepackage{amsfonts}    
\usepackage{dsfont}
\usepackage{pdfpages}
\usepackage{verbatim}
\usepackage{tensor}
\usepackage{mathrsfs}
\usepackage[mathscr]{euscript}
\usepackage{tikz}
\usetikzlibrary{hobby}

\newcommand{\ba}{\begin{align}}

\newcommand{\be}{\begin{equation}}
\newcommand{\ee}{\end{equation}}
\def\bd{\begin{tikzpicture}}
\def\ed{\end{tikzpicture}}

\allowdisplaybreaks[1]

\title{The physical states of the Hybrid Formalism}

\author{Matthias R.\ Gaberdiel}
\author{and Kiarash Naderi} 
\affiliation{Institut f\"ur Theoretische Physik, ETH Zurich, \\
\hspace*{0.3cm} 8093 Z\"urich, Switzerland}
\emailAdd{gaberdiel@itp.phys.ethz.ch}
\emailAdd{knaderi@phys.ethz.ch}

\abstract{String theory on ${\rm AdS}_3\times {\rm S}^3 \times\mathbb{T}^4$ with one unit ($k=1$) of NS-NS flux is considered in the hybrid formalism of Berkovits, Vafa \& Witten (BVW). Using the free field realisation of the world-sheet theory at $k=1$, we identify explicitly the BRST cohomology classes corresponding to some of the low-lying states of the dual CFT. In particular, we do this for the ${\cal N}=4$ superconformal generators of the symmetric orbifold theory, and we confirm these identifications by showing that the worldsheet correlators reproduce the expected dual CFT answer. Along the way we note that the physical vertex operators on the worldsheet have a simpler form if one works with a different, but equivalent, choice for the BRST operators relative to BVW.}

\begin{document}

\maketitle

\section{Introduction}

The AdS/CFT duality for the case of AdS$_3$/CFT$_2$ has recently been understood in detail. In particular, it was shown in \cite{Eberhardt:2018ouy}  that string theory on $\text{AdS}_3 \times \text{S}^3 \times \mathbb{T}^4$ with $k=1$ units of NS-NS flux has exactly the same spacetime spectrum as the symmetric orbifold of $\mathbb{T}^4$ in the large $N$ limit. (Earlier work on the $k=1$ theory includes \cite{Gaberdiel:2018rqv,Giveon:2005mi,Giribet:2018ada,Gaberdiel:2017oqg,Ferreira:2017pgt}.) Furthermore,  the correlators of the symmetric orbifold theory are correctly reproduced by this worldsheet theory \cite{Eberhardt:2019ywk}, see also \cite{Eberhardt:2020akk,Dei:2020zui,Knighton:2020kuh} for further developments.

While the definition of the $k=1$ theory in the RNS formalism is a bit delicate, the construction of the worldsheet theory in the hybrid formalism of \cite{Berkovits:1999im} is unproblematic. In this approach  the theory consists of a $\mathfrak{psu}(1,1|2)_1$ WZW model, a topologically twisted sigma-model on $\mathbb{T}^4$, as well as a $(\rho,\sigma)$ ghost system, see  Appendix~\ref{appendix:superconformal-algebra-hybrid} for more details. In particular, the $\mathfrak{psu}(1,1|2)_1$ theory is well-defined, and it actually exhibits a free field realisation in terms of symplectic bosons and fermions that seems to generalise naturally to the higher dimensional case \cite{Gaberdiel:2021iil,Gaberdiel:2021jrv}. 
\smallskip

The above duality has so far been established and tested at the `free' point, i.e.\ where the dual CFT is the actual symmetric orbifold theory. In order to understand the duality more deeply, it would be important to study the deformation away from this special point, in particular, given the fact that the pure NS-NS background for AdS$_3$ is somewhat special. As a preparation for this analysis we need to develop a better grasp for how the different states of the symmetric orbifold theory are realised on the worldsheet. This is not as straightforward as it may seem since the characterisation of the physical states in the hybrid theory is via some rather complicated BRST cohomology, and only a few states have been worked out in detail so far \cite{Troost:2011fd,Gaberdiel:2011vf,Gerigk:2012cq,gerick:thesis}. 

In this paper we begin a systematic study of this BRST cohomology, and identify explicitly some of the low-lying states of the dual symmetric orbifold. In particular, we do this for the ${\cal N}=4$ generators and the free fermions and bosons of the spacetime symmetric orbifold of $\mathbb{T}^4$.   (We will also identify another spin-$2$ current, namely the Sugawara tensor associated to the $\mathfrak{su}(2)_1$ algebra of the spacetime $\mathbb{T}^4$ theory.)
We then confirm these identifications by checking that the corresponding correlation functions reproduce the symmetric orbifold answer; this generalises, among other things, the analysis of \cite{Bertle:2020sgd} to the hybrid formalism. 

Apart from these detailed (and successful) checks, there are two structural insights that emerge. First, we verify that the $\mathfrak{su}(2)$ $R$-symmetry of the symmetric orbifold theory can indeed be identified with the global $\mathfrak{su}(2)$ subalgebra of the $\mathfrak{psu}(1,1|2)_1$ symmetry on the worldsheet, see eq.~(\ref{su2R}); while this is certainly what one should have expected on abstract grounds, it was not clear, e.g.\ from the analysis of \cite{Eberhardt:2018ouy},  how this would work out in detail. The other structural insight concerns the explicit form of the BRST operators. We find that the BRST analysis is significantly simplified --- e.g. physical states can be taken to have definite ghost number ---  if one works with a slightly different, but equivalent, version of the BRST operators, where the final similarity transformation of \cite{Berkovits:1999im}, see the discussion around eq.~(4.14) there, has not been performed. 
\medskip

The paper is organised as follows. In Section~\ref{section:hybrid-free}, see also Appendix~\ref{appendix:superconformal-algebra-hybrid}, we explain our conventions and set up the notation. The low-lying physical states are identified in Section~\ref{section:states}. In particular, this is done for the spacetime ${\cal N}=4$ fields in Section~\ref{sec:3.1}, and then for the free bosons and fermions in Section~\ref{sec:3.2}. In Section~\ref{sec:correlators} we determine the correlators of these vertex operators on the worldsheet, and confirm that they reproduce the corresponding spacetime amplitudes. Section~\ref{sec:conclusions} contains our conclusions, and there are a number of appendices that are referred to throughout the main text.

\section{The hybrid formalism}
\label{section:hybrid-free}

The worldsheet theory corresponding to string theory on $\text{AdS}_3 \times \text{S}^3 \times \mathbb{T}^4$ with one $k=1$ unit of NS-NS flux consists, in the hybrid formalism of \cite{Berkovits:1999im}, of a $\mathfrak{psu}(1,1|2)_1$ WZW model, a topologically twisted sigma-model on $\mathbb{T}^4$, as well as a $(\rho,\sigma)$ ghost system. The worldsheet has an ${\cal N}=4$ superconformal symmetry, and the physical states $\psi$ of the string theory are characterised by the BRST cohomology,\footnote{Our conventions differ by a factor of $2$ for the normalisation of $J$ relative to \cite{Berkovits:1999im}, see Appendix~\ref{appendix:superconformal-algebra-hybrid}. We also consider the generators without the similarity transformation of eq.~(4.14) of \cite{Berkovits:1999im}, see Appendix~\ref{appendix:superconformal-algebra-hybrid} for more details.} 
\begin{equation} \label{eq:physical-conditions}
G^+_0 \psi = \tilde{G}^+_0 \psi = (J_0- \tfrac{1}{2}) \psi= T_0 \psi = 0  \ , \qquad \psi \sim \psi + G^+_0 \tilde{G}^+_0 \psi^{\prime} \ . 
\end{equation}
For the case at hand, the $\mathfrak{psu}(1,1|2)_1$ algebra has a free field realisation in terms of symplectic bosons and complex fermions, see Appendix~\ref{appendix:free-field}, and as a consequence, the ${\cal N}=4$ superconformal generators that appear in (\ref{eq:physical-conditions}) take a relatively simple form; this is described in detail in Appendix~\ref{appendix:superconformal-algebra-hybrid}. There is one subtlety in that the  free field realisation actually leads to $\mathfrak{u}(1,1|2)_1$, and in order to reduce this to $\mathfrak{psu}(1,1|2)_1$ one has to impose in addition that \cite{Dei:2020zui}
\be\label{eq:coset}
Z_n \psi = 0 \qquad n\geq 0 \ . 
\ee
It is one of the aims of this paper to find various low-lying physical states of the worldsheet theory, i.e.\  states satisfying (\ref{eq:physical-conditions}) and (\ref{eq:coset}). In particular, we shall identify the states that correspond to the generators of the ${\cal N}=4$ superconformal symmetry of the dual CFT (the symmetric orbifold of $\mathbb{T}^4$). We shall also find the worldsheet description of the chiral fermion and boson fields of this symmetric orbifold. One important lesson of our analysis is that the $R$-symmetry $\mathfrak{su}(2)$ of the spacetime \linebreak ${\cal N}=4$ algebra turns out to be `identified' with the natural $\mathfrak{su}(2) \subset \mathfrak{psu}(1,1|2)$ worldsheet symmetry. More precisely, our physical worldsheet states transform under the action of the $\mathfrak{su}(2) \subset \mathfrak{psu}(1,1|2)$ zero modes exactly as the corresponding spacetime fields transform under the spacetime $R$-symmetry. While this is certainly what one should have expected, it was not clear how this would come about, since the spacetime $R$-symmetry acts on the fermions of the symmetric torus orbifold, while the worldsheet $\mathfrak{su}(2)$ symmetry is associated to the ${\rm S}^3$ factor, and hence does not act on the  torus fermions on the worldsheet. 
\medskip

In order to confirm that we have correctly identified these states we shall then, in a second step, calculate their correlation functions using the techniques developed in  \cite{Dei:2020zui}. In particular, we shall consider the $n$-point functions 
\begin{equation} \label{eq:correlators-on-s2}
\left< \Big(\prod_{i=1}^{n-2} W(u_i)\Big)\, [(e^{i\sigma})_{0}(e^{-\rho-i H} \Phi_1)] \, \Phi_2 \, (\tilde{G}^-_{-1} \Phi_3) \cdots (\tilde{G}^-_{-1} \Phi_n)\right> \ , 
\end{equation}
where we have used eqs.~(3.2) and (3.8) of \cite{Dei:2020zui}, and replaced  $(G^-)^{-1}\sim e^{i\sigma}$ and $(\tilde{G}^+)^{-1} \sim e^{-\rho-i H}$. Moreover, $W$ is the vertex operator associated to the ``vacuum" state $\left| 0 \right>^{(1)}$ with $Y_0=2$ (and $Z_0=0$), see \cite{Dei:2020zui} for more details. We assume here that the ghost picture numbers of the vertex operators $\Phi_i$ sum to  $-2n$.  Actually, we shall only consider $2$- and $3$-point functions since this will be sufficient to extract the relevant OPE coefficients. 

The structure of (\ref{eq:correlators-on-s2}) suggests that there is a direct relation between the ghost picture number and the $Y_0$ eigenvalue of each operator, and indeed, one can choose the physical vertex operators such that they satisfy 
\begin{equation} \label{eq:v0-ghost}
P = - 2 - y \ , 
\end{equation}
where $P$ and $y$ are the ghost picture number and $Y_0$ eigenvalue, respectively. This is what we should have expected since the condition that the ghost picture numbers add up to $-2n$ is then equivalent to 
\be
\sum_{i=1}^{n} y_i = 0 \ . 
\ee
On the other hand,  since each $\tilde{G}^-_{-1}$ operator reduces $Y_0$ by $-2$,  the total $Y_0$ charge of all the vertex operators in the correlator is 
\be
(n-2)\cdot  2 + \sum_{i=1}^{n} y_i + (n-2) \cdot (-2) = 0 \ , 
\ee
and hence vanishes (as required) provided that (\ref{eq:v0-ghost}) holds.

\section{The physical states on the worldsheet}
\label{section:states}

In this section we find the worldsheet states corresponding to certain low-lying states of the 
symmetric orbifold theory. We shall first concentrate on the ${\cal N}=4$ superconformal  fields of  the symmetric orbifold theory; they are `compactification independent' since they would also, say, be present if we considered instead of ${\rm AdS}_3 \times {\rm S}^3 \times \mathbb{T}^4$ the background ${\rm AdS}_3 \times {\rm S}^3 \times {\rm K3}$. In the second half of this section (see Section~\ref{sec:3.2}) we shall then also construct the states corresponding to the $4$ free bosons and fermions of the symmetric orbifold of $\mathbb{T}^4$; they are obviously `compactification dependent' since, say, the K3 theory does not have such fields. 

\subsection{The spacetime ${\cal N}=4$ fields}\label{sec:3.1}

We begin our analysis with finding the worldsheet states associated to the generators of the  $\mathcal{N}=4$ superconformal algebra  (see Appendix~\ref{appendix:t4} for our conventions).

\subsubsection{The $\mathfrak{su}(2)_1$ generators}
\label{section:su2}

The states of lowest spacetime conformal dimension ($h=1$) are the $R$-symmetry $\mathfrak{su}(2)$ currents --- in the convention of Appendix~\ref{appendix:t4} they are  $J^{++}$, $J$ and $J^{--}$  ---  and we shall first aim to identify the corresponding worldsheet states. Following \cite{Gerigk:2012cq,gerick:thesis} we will make the ansatz 
\begin{equation} \label{eq:comp-indep}
\Phi^a \, e^{2\rho+i\sigma+iH} \ , 
\end{equation}
where we write $\Phi^a=[\phi^a]^{\sigma}$ (with $a=++$, $3$, $--$) since these fields come from the untwisted sector of the symmetric orbifold, and hence arise in the first spectrally flowed sector on the worldsheet. (Here states in the $\sigma$-flowed sector are labelled by $[\phi^a]^{\sigma}$, where $\phi^a$ is a state in the usual Ramond sector representation, and the bracket $[ \cdot ]^{\sigma}$ means that the action of the modes is twisted by $\sigma$, see eq.~(\ref{eq:spectral-flow-representation}) for more details.) Because these states are `compactification independent' we expect that $\Phi^a$ is a state in the $\mathfrak{psu}(1,1|2)_1$ theory, i.e.\ does not involve the topologically twisted $\mathbb{T}^4$ factor on the worldsheet. We also note that the vertex operator in eq.~(\ref{eq:comp-indep}) has ghost picture $P=-2$.

The physical state conditions (\ref{eq:physical-conditions}) are satisfied provided that  
\begin{equation} \label{eq:comp-physical}
L_n \Phi^a=Q_n \Phi^a=0 \ , \quad  (n \geq 0) \ ,
\end{equation}
where $L_n$ are the Virasoro generators of  $\mathfrak{psu}(1,1|2)_1$, while $Q_n$ is defined as in \cite{Dei:2020zui}, see also eq.~(\ref{eq:q}) in Appendix~\ref{appendix:superconformal-algebra-hybrid},
\begin{equation} \label{eq:qn}
Q_n = 2 \sum_{r,s,p,q} (r-s)\xi^+_r \xi^-_s \chi^+_p \chi^-_q \delta_{r+s+p+q,n} \ .
\end{equation}
We are looking for states with spacetime conformal dimension $h=1$, i.e.\ with $J^3_0$ eigenvalue $J^3_0=1$ on the worldsheet. Let us denote by $N$, $n$ and $m$ the $L_0$, $J^3_0$ and $K^3_0$ eigenvalues before spectral flow, respectively. Since spectral flow by one unit shifts the $J^3_0$ eigenvalue by $+\frac{1}{2}$, see eq.~(\ref{psu-j3-sigma}), we need $n=\frac{1}{2}$. The mass-shell ($L_0=0$) condition then implies that 
\be\label{mass-shell}
N  = (\tfrac{1}{2} - m ) \ . 
\ee
Since $N\in \mathbb{N}_0$, $m$ must be half-odd-integer with $m\leq \frac{1}{2}$. We can put more constraints on the possible values of $m$ by noting that if the $K^3_0$ eigenvalue equals $m$, we need at least $(1-2m)$ fermionic generators $F^-$ before spectral flow, where $F$ is either a mode of $\chi$ or one of $\psi$. Because of the Fermi statistics of these fermionic modes it then follows that 
\begin{equation}
N=(\tfrac{1}{2} - m ) \geq m^2 - \tfrac{1}{4} \ ,
\end{equation}
which together with $m \leq \tfrac{1}{2}$ implies $m \geq -\tfrac{3}{2}$, or equivalently $m=\frac{1}{2},-\frac{1}{2},-\frac{3}{2}$. This leaves us, as expected, with the $K^3_0$ eigenvalues of the three $\mathfrak{su}(2)$ currents that are $0, \pm 1$. 

Let us discuss the three cases in turn. For $m=\frac{1}{2}$, eq.~(\ref{mass-shell}) implies that $N=0$, i.e.\ that the state before spectral flow is a Ramond ground state. $K^3_0=\frac{1}{2}$ then implies that it also does not have any fermionic zero mode excitations, and $Z_0=0$ together with $J^3_0=\frac{1}{2}$ require that it must be the state\footnote{The normalisations have been fixed for future convenience.}
\begin{equation}
\phi^{++}=-\left|\tfrac{1}{2},0\right> \ . 
\end{equation}
It is straightforward to check that $\Phi^{++}=[\phi^{++}]^{\sigma}$ satisfies indeed the physical state conditions  (\ref{eq:physical-conditions}) and (\ref{eq:coset}). For the following it will be convenient to rewrite this state in a somewhat more suggestive form. Let us introduce the states
\be \label{eq:omegam1m2}
\Omega_{m_1,m_2} =\left[\chi_0^- \psi_0^- \left|m_1,m_2\right>\right]^{\sigma}\ 
\ee
that are affine highest weight with respect to the $\mathfrak{su}(2)_1$ subalgebra of $\mathfrak{psu}(1,1|2)_1$. Then we can write 
\be
\Phi^{++}  = - \bigl[ \, \left|\tfrac{1}{2},0\right> \bigr]^\sigma =  K^+_{-1} \Omega_{\frac{1}{2},0} \ . 
\ee

For $m=-\frac{1}{2}$ we have $N=1$, i.e.\ we need one non-zero excitation mode, and (at least) two fermionic $\psi^-$ or $\chi^-$ modes. We make the schematic ansatz 
\begin{equation}
\sum_{S} S^{\pm}_{-1} \chi^-_0 \psi^-_0 \left| m_1,m_2\right>  +  \sum_{F} F^-_{-1} F^-_{0} \left| m_1,m_2 \right> \ ,
\end{equation} 
where $S$ is either $\xi$ or $\eta$, and $F$ is either $\chi$ or $\psi$. After imposing the physical state conditions (\ref{eq:comp-physical}) we find that the only physical states are\footnote{This calculation as well as the calculations below were done with the help of the Virasoro Mathematica  package \cite{mathematica}.}
\begin{equation}
(\chi^{-}_{-1} \psi^-_0 + \chi^-_0 \psi^-_{-1}) \left| \tfrac{1}{2} , 0 \right> \ , \qquad
 \eta^+_{-1} \chi^-_0 \psi^-_0 \left| 0,0\right> \ , \qquad 
 \chi^-_{-1} \chi^-_{0} \left| 0,\tfrac{1}{2} \right> \ .
\end{equation}
The middle state $[\eta^+_{-1} \chi^-_0 \psi^-_0 \left| 0,0\right>]^{\sigma}=J^+_0 [\chi^-_0 \psi^-_0\left|0,-\frac{1}{2}\right>]^{\sigma}$ is actually an $\mathfrak{su}(2)$ singlet (and hence not what we expect); in any case, we can show that it is BRST exact since $J^+_0$ corresponds to the $L_{-1}$ generator in the dual CFT, and $\Omega_{0,-\frac{1}{2}} = [\chi^-_0 \psi^-_0\left|0,-\frac{1}{2}\right>]^{\sigma}$ describes the vacuum of the dual CFT, see Appendix~A of \cite{Dei:2020zui}; using the results of \cite{Dei:2019osr} this then implies that the state must be BRST exact. We have also checked explicitly that it decouples from the correlation functions to be discussed below in Section~\ref{sec:correlators}. The last state $[\chi^-_{-1} \chi^-_{0} \left| 0,\frac{1}{2} \right>]^{\sigma}$ (which is again an $\mathfrak{su}(2)$ singlet) is  also BRST exact, since it can be written as 
\begin{equation}
\Big[\chi^-_{-1} \chi^-_{0} \left| 0,\tfrac{1}{2} \right>\Big]^{\sigma}=Q_0\Big[\tfrac{1}{2} \chi^-_0 \psi^-_{-1} \left|\tfrac{1}{2},0\right>\Big]^{\sigma} \ , 
\end{equation}
and this implies BRST exactness according to the analysis of \cite{gerick:thesis}. The only physical state that remains is therefore
\begin{equation}
\phi^{3}=-\tfrac{1}{2}(\chi^{-}_{-1} \psi^-_0 + \chi^-_0 \psi^-_{-1}) \left| \tfrac{1}{2} , 0 \right> = - \tfrac{1}{2} K^-_{-1} \, \phi^{++} \ , 
\end{equation}
where the last identity implies that
\be\label{Phi3}
\Phi^3 = [ \phi^3 ]^{\sigma} = - \tfrac{1}{2} K^-_0 \Phi^{++}  =K^3_{-1} \Omega_{\frac{1}{2},0} \ .
\ee

The analysis for $m=-\frac{3}{2}$ is similar albeit a bit more complicated since now $N=2$, and hence we need a level $2$ descendant before spectral flow.  Apart from BRST exact states we have found one physical state 
\begin{equation}
\phi^{--}=-\chi^{-}_{-1} \psi^-_{-1} \chi^-_0 \psi^-_{0} \left| \tfrac{1}{2} , 0 \right> = - \tfrac{1}{2} K^-_{-1} K^-_{-1} \phi^{++} \ , 
\end{equation}
and the last equation demonstrates again that the state is physical, i.e.\ satisfies  (\ref{eq:physical-conditions}) and (\ref{eq:coset}). We also note that in analogy to (\ref{Phi3}) we can write it as
\be
\Phi^{--} = [ \phi^{--}]^{\sigma} = - \tfrac{1}{2}\, K^-_0 K^-_0\, \Phi^{++}  = K^-_{-1} \Omega_{\frac{1}{2},0} \ . 
\ee
Putting back again the ghost-fields we have therefore found candidates for the spacetime $R$-symmetry currents, namely 
\begin{subequations} \label{su2R}
\begin{align} 
\textbf{J}^{++} & =\left[\, -\left| \tfrac{1}{2} , 0 \right> \right]^{\sigma} e^{2\rho + i \sigma +i H}  \\[2pt] 
\textbf{J}^{3} &= - \tfrac{1}{2} K^{-}_{0}\left[ \, - \left| \tfrac{1}{2} , 0 \right> \right]^{\sigma}e^{2\rho + i \sigma +i H}=\tfrac{1}{2}\left[ K^{-}_{-1} \left| \tfrac{1}{2} , 0 \right> \right]^{\sigma} e^{2\rho + i \sigma +i H} \\[2pt]
\textbf{J}^{--} & =-\tfrac{1}{2} K^{-}_{0} K^{-}_{0} \left[ - \, \left| \tfrac{1}{2} , 0 \right> \right]^{\sigma} e^{2\rho + i \sigma +i H} = \tfrac{1}{2}\left[ K^{-}_{-1} K^{-}_{-1} \left| \tfrac{1}{2} , 0 \right> \right]^{\sigma} e^{2\rho + i \sigma +i H} \ .
\end{align}
\end{subequations}
From the way we have written these vertex operators it is manifest that they transform in the triplet representation with respect to $\mathfrak{su}(2) \subset \mathfrak{psu}(1,1|2)$. While this was to be expected on general grounds, it is very satisfying to see it being realised in our explicit construction.

\subsubsection{Stress energy tensor}
\label{section:t}

Next we want to identify the worldsheet state corresponding to the stress energy tensor of the dual symmetric orbifold theory. Again we make the ansatz 
\begin{equation}\label{compind}
\Phi \, e^{2\rho+i\sigma+iH} \ , 
\end{equation}
where $\Phi=[\Phi_T]^{\sigma}$ should be a vertex operator in $\mathfrak{psu}(1,1|2)_1$ that transforms as a singlet with respect to $\mathfrak{su}(2) \subset \mathfrak{psu}(1,1|2)$;  
in particular, thus $K^3_0=-\frac{1}{2}$ before spectral flow. The spacetime conformal dimension of $\Phi$ should be $2$, i.e.\ it should have $J^3_0$ eigenvalue $+\frac{3}{2}$ before spectral flow, and it therefore follows from eq.~(\ref{spectral-psu-ln}) that the excitation number equals $N=2$ before spectral flow. The most general ansatz for $\Phi_T$ is then a linear combination of the states 
\begin{subequations} \label{eq:t-general-states}
\begin{gather}
S^{\pm}_{-2} \chi^-_0 \psi^-_0 \left| m_1,m_2 \right>\ , \ \  S^{\pm}_{-1} S^{\pm}_{-1} \chi^-_0 \psi^-_0 \left| m_1,m_2 \right>\ , \ \ S^{\pm}_{-1} F^-_{-1} F^-_0 \left| m_1,m_2 \right> \ , \\ F^-_{-2} F^-_{0} \left| m_1,m_2 \right>\ ,  \ \ \chi^-_{-1} \psi^-_{-1} \left| m_1,m_2 \right>\ ,  \ \ F^+_{-1} F^-_{-1} \chi^-_0 \psi^-_0 \left| m_1,m_2 \right> \ , 
\end{gather}
\end{subequations}
where $S$ is either $\xi$ or $\eta$, and $F$ is either $\chi$ or $\psi$. We then impose the physical state conditions, i.e.\ eqs.~(\ref{eq:comp-physical}) and (\ref{eq:coset}); we also require that the state is an $\mathfrak{su}(2)$ singlet (i.e.\ that it satisfies $K^\pm_0\Phi=0$), as well as quasiprimary in spacetime (i.e.\ that it satisfies $J^-_0 \Phi=0$). Then there are, up to BRST exact states, two linear combinations. The linear combination that corresponds to the spacetime stress energy tensor turns out to be 
\begin{align} \label{eq:t}
\Phi_T = & \alpha \Big( -3 \xi^+_{-2} \chi_0^- \psi_0^- \left| 1, 0 \right>
-4 \, \xi^-_{-2} \chi_0^- \psi_0^- \left| \tfrac{3}{2}, \tfrac{1}{2} \right>
+3\, \eta^+_{-2} \chi_0^- \psi_0^- \left| \tfrac{1}{2}, \tfrac{1}{2} \right>
+\eta^-_{-2} \chi_0^- \psi_0^- \left| 1, 1 \right> \nonumber \\
& \quad +3\, \xi^+_{-1} \eta^-_{-1} \chi_0^- \psi_0^- \left| 1, \tfrac{1}{2} \right>
+3 \, \xi^-_{-1} \eta^+_{-1} \chi_0^- \psi_0^- \left| 1, \tfrac{1}{2} \right>
-6 \, \xi^+_{-1} \xi^-_{-1} \chi_0^- \psi_0^- \left| \tfrac{3}{2}, 0 \right>\\
& \quad -6 \, \xi^-_{-1} \xi^-_{-1} \chi_0^- \psi_0^- \left| 2, \tfrac{1}{2} \right>
+6 \, \xi^+_{-1} \eta^+_{-1} \chi_0^- \psi_0^- \left| \tfrac{1}{2}, 0 \right>
+2\,  \xi^-_{-1} \eta^-_{-1} \chi_0^- \psi_0^- \left| \tfrac{3}{2}, 1 \right>\Big) \ .  \nonumber 
\end{align}
From the calculation of the correlators in Section~\ref{sec:correlators} we shall see that the correct normalisation is $\alpha=\frac{1}{2}$, see eq.~(\ref{eq:normalization-of-t}) below. 
The other linearly independent physical state can be taken to be 
\begin{align} \label{eq:sugawara-su2}
\Phi_{T_{\mathfrak{su(2)}}} = & \beta \Big[(\chi^-_{-2} \psi_0^- +2 \, \chi^-_{-1} \psi_{-1}^- +\chi^-_{0} \psi_{-2}^- - \chi^+_{-1} \psi^-_{-1} \chi_0^- \psi_0^- - \chi^-_{-1} \psi^+_{-1} \chi_0^- \psi_0^-) \left|1,\tfrac{1}{2} \right> \nonumber \\
& \quad +2 \, \xi^-_{-2} \chi_0^- \psi_0^- \left| \tfrac{3}{2}, \tfrac{1}{2} \right>
+2 \, \xi^+_{-2} \chi_0^- \psi_0^- \left| 1, 0 \right>
+2 \, \xi^-_{-1} \xi^-_{-1} \chi_0^- \psi_0^- \left| 2, \tfrac{1}{2} \right> \\
& \quad +2 \, \xi^+_{-1} \xi^-_{-1} \chi_0^- \psi_0^- \left| \tfrac{3}{2}, 0 \right> \Big] \nonumber \ ,
\end{align}
where the state in the first line equals\footnote{The remaining terms in eq.~(\ref{eq:sugawara-su2}) are needed to make the state BRST closed; however, these terms are by themselves not BRST exact.}
\be \label{eq:sugarawa}
\tfrac{4}{3} \big[  K^3_{-1} K^3_{-1}+\tfrac{1}{2} (K^+_{-1} K^-_{-1} + K^-_{-1} K^+_{-1})\big] \Omega_{1,\frac{1}{2}} \ .
\ee
This suggests that the state in (\ref{eq:sugawara-su2})  will be associated to the Sugawara stress energy tensor of the spacetime R-symmetry; this viewpoint then also indicates that we should set $\beta=\frac{1}{4}$,  and this will be confirmed by the calculations of Section~\ref{sec:tsu2}. 

\subsubsection{Supercurrents}
\label{section:supercurrents}

The spacetime theory also has two pairs of supercurrents $G^{\pm}$ and $\tilde{G}^{\pm}$  that transform as doublets with respect to the spacetime $R$-symmetry, 
\begin{equation}
\begin{pmatrix}
G^+ \\ \tilde{G}^-
\end{pmatrix} \qquad \hbox{and} \qquad 
\begin{pmatrix}
\tilde{G}^+ \\ G^-
\end{pmatrix} \ ,
\end{equation}
see Appendix~\ref{appendix:t4}. For the first doublet, $(G^+,\tilde{G}^-)$,  we make the ansatz 
\begin{align} \label{eq:g+}
G^+ & =  \phi_+ \, e^{-\rho+i \sigma} \tilde{G}^-_C \ ,  \\ \label{eq:gtilde-}
\tilde{G}^- & = \phi_- \, e^{-\rho+i \sigma}  \tilde{G}^-_C \ ,
\end{align}
where $\tilde{G}^-_C$ is defined in eq.~(\ref{eq:tildeGC}). This ansatz has ghost picture $P=1$.\footnote{While our ansatz in (\ref{eq:g+}) and (\ref{eq:gtilde-}) is not of the `compactification-independent' form of eq.~(\ref{compind}), it can be obtained from the corresponding compactification-independent states
upon applying $P_+$ twice. The other two supercurrents, see eqs.~(\ref{eq:tildeG+ansatz1}) and (\ref{eq:g-}), on the other hand, do not have  a `compactification-independent' description, i.e.\ cannot be brought into the form of eq.~(\ref{compind}).}  The physical state conditions (\ref{eq:physical-conditions}) are then satisfied provided that
\begin{equation} \label{eq:supercurrents-conditions-1}
L_0 \phi_{\pm} = \phi_{\pm}\ , \qquad  L_n \phi_{\pm}=0  \ \ (n \geq 1)\ , \qquad Q_{n} \phi_{\pm}=0 \ \ (n \geq -3) \ , 
\end{equation}
and the unique solutions are (note that $\tilde{G}^-=-J^{--}_0 G^+$ in our conventions),\footnote{We have only been able to show uniqueness in the $P=-1$ picture; the states in (\ref{eq:g+}) and (\ref{eq:gtilde-}) are then the images of these unique states under $P_+$.}
\be \label{eq:g+-gtilde-}
\phi_+  =  \left[ \chi^+_{-1} \chi^-_{-1} \chi^-_{0} \left| 0, 1 \right>\right]^{\sigma} \ , \qquad \qquad
\phi_-  = -\left[ \chi^-_{-2} \chi^-_{-1} \chi^-_{0} \left| 0, 1 \right>\right]^{\sigma}  = 
- K^-_0 \phi_+ \ . 
\ee
In particular, the second expression makes it clear that these fields transform in the doublet representation w.r.t\ the $\mathfrak{su}(2)$ zero mode subalgebra of $\mathfrak{psu}(1,1|2)_1$. 
\smallskip

\noindent For the other doublet, i.e.\ $\tilde{G}^{+}$ and $G^{-}$, we make the ansatz
\begin{align}\label{eq:tildeG+ansatz1}
\tilde{G}^+ & = 
\phi_{+} e^{5\rho+i\sigma+2iH} \tilde{G}^{+}_C   \ , \\ \label{eq:g-}
{G}^- & = 
\phi_{-} e^{5\rho+i\sigma+2iH} \tilde{G}^{+}_C  \ ,
\end{align}
where $\tilde{G}^+_C$ is defined in eq.~(\ref{eq:tildeGC}). This ansatz has ghost picture $P=-5$. The physical state conditions (\ref{eq:physical-conditions}) are satisfied if 
\begin{equation} \label{eq:supercurrents-conditions}
L_0 \phi_{\pm} = \phi_{\pm}\ , \qquad  L_n \phi_{\pm}=0  \ \ (n \geq 1)\ , \qquad Q_{n} \phi_{\pm}=0 \ \ (n \geq 3) \ .
\end{equation}
The analysis of the physical state conditions in this case are actually quite complicated, and it is easier to move to the $-3$ picture by acting by $P_+$ on the ansatz for $\tilde{G}^+$ and $G^-$, which should map physical states to physical states
\begin{align}\label{eq:tildeG+ansatz}
\tilde{G}^+ & \rightarrow 
\chi_{+} e^{3\rho+i\sigma+iH} \tilde{G}^{+}_C -2\phi_{+} e^{4\rho+i\sigma+2i H} \ , \\ \label{eq:G-ansatz}
{G}^- & \rightarrow
\chi_{-} e^{3\rho+i\sigma+iH} \tilde{G}^{+}_C -2\phi_{-} e^{4\rho+i\sigma+2i H} \ , 
\end{align}
where $Q_2 \phi_{\pm}=- \chi_{\pm}$. We note that the vertex operators in eqs.~(\ref{eq:tildeG+ansatz}) and (\ref{eq:G-ansatz}) do not have definite ghost numbers, see eq.~(\ref{ghost-calc}). This is a reflection of the fact that also $G^+$ in  (\ref{Gp}) does not have a definite ghost number.

With this ansatz, the physical state conditions (\ref{eq:physical-conditions}) are satisfied provided that 
\begin{subequations} \label{eq:supercurrents-conditions-p=-3}
	\begin{align}
	L_0 \chi_{\pm} & = -\chi_{\pm} \ , \qquad \qquad \ \,  L_0 \phi_{\pm}=\phi_{\pm} \ , \\
	L_n \chi_{\pm} & = 0 \ \ (n \geq 1) \ , \qquad L_n \phi_{\pm}=0 \ \ (n \geq 1) \ , \\ 
	Q_n \chi_{\pm} & =0 \ \ (n \geq 1)\ , \qquad  Q_n \phi_{\pm}=0 \ \ (n \geq 3) \ .
	\end{align}
\end{subequations}
Up to BRST exact states the unique solution (with spacetime conformal dimension $h=\frac{3}{2}$) is then 
\begin{equation} \label{eq:supercurrents-gtilde+-phi}
\chi_+ = - 2 \, \left[ \psi^-_0 \left| 1, 0 \right>\right]^{\sigma} \ , \qquad \qquad 
\phi_+ = \left[ \psi^+_{-1} \psi^-_{-1} \psi^-_0 \left| \tfrac{3}{2}, -\tfrac{1}{2} \right>\right]^{\sigma} \ , 
\end{equation}
and ($G^-=J^{--}_0 \tilde{G}^+$ in our conventions)
\begin{equation} \label{eq:supercurrents-g--phi}
\chi_- =  - 2\, \left[ \psi^-_{-1} \chi^-_0 \psi^-_{0} \left| 1,0 \right>\right]^{\sigma}  = K^-_0 \chi_+
\ , \qquad 
\phi_- =  \left[ \psi^-_{-2} \psi^-_{-1} \psi^-_0 \left| \tfrac{3}{2}, -\tfrac{1}{2} \right>\right]^{\sigma}  =  K^-_0 \phi_+\ . 
\end{equation}
Again, this shows that these fields transform correctly (namely as a doublet) with respect to the $\mathfrak{su}(2)$ zero mode subalgebra of $\mathfrak{psu}(1,1|2)_1$.

\subsection{The free bosons and fermions of $\mathbb{T}^4$}\label{sec:3.2}
In this subsection, we derive the worldsheet states that are dual to the free fermions and bosons of the symmetric orbifold of $\mathbb{T}^4$.

\subsubsection{Free Fermions}
\label{section:fermions}
The $4$ free fermions of the spacetime theory transform as two doublets with respect to the spacetime $R$-symmetry, namely
\begin{equation} \label{eq:fermions-doublet-f}
\begin{pmatrix}
\psi^1 \\
\bar{\psi}^2
\end{pmatrix} \qquad \hbox{and} \qquad 
\begin{pmatrix}
\psi^2 \\
-\bar{\psi}^1
\end{pmatrix} \ .
\end{equation}
We make the following ansatz\footnote{See also Section 5.3.2 of \cite{gerick:thesis} where it is argued that the states of the form (\ref{eq:fermions-states}) are part of the spectrum for superstrings on $\text{AdS}_3 \times \text{S}^3 \times \mathbb{T}^4$.}
\begin{equation} \label{eq:fermions-states}
\Psi^j_{\pm}=\phi_{\pm} e^{\rho+i \sigma} e^{i H^j} \ .
\end{equation}
We note that this ansatz has ghost picture $P=-1$. The physical state conditions are
\begin{equation} \label{eq:psi-condition}
L_n \phi_{\pm}=0  \ \ (n \geq 0)\ , \qquad Q_{n} \phi_{\pm}=0 \ \ (n \geq -1) \ . 
\end{equation}
Up to BRST exact states, we find the following unique solutions (with spacetime conformal dimension $h=\tfrac{1}{2}$)
\begin{equation} \label{eq:fermions-+}
\phi_+=[\chi^-_{0}\left|0,0\right>]^{\sigma} \ , \qquad \qquad \phi_-=[ \chi^{-}_{-1} \chi^{-}_{0} \psi^{-}_{0} \left|0,0\right>]^{\sigma}=K^-_0 \phi_+ \ .
\end{equation}
Therefore, with respect to the $\mathfrak{su}(2)$ zero mode subalgebra of $\mathfrak{psu}(1,1|2)_1$ they form doublets
\begin{equation} \label{eq:wfermions-doublet-f}
\begin{pmatrix}
\Psi^1_{+} \\
\Psi^1_{-}
\end{pmatrix} \qquad \hbox{and} \qquad 
\begin{pmatrix}
\Psi^2_{+} \\
\Psi^2_{-}
\end{pmatrix} \ .
\end{equation}
For the correlators in the next section, it is convenient to express these states also in a different ghost picture, namely in $P=-3$
\begin{equation} \label{eq:fermions-p=-3}
\tilde{\Psi}_{\pm}^j=\tilde{\phi}_{\pm} e^{3\rho+i \sigma +iH} e^{i H^j} \ ,
\end{equation}
where
\begin{equation} \label{eq:fermions-p-+}
\tilde{\phi}_+ = -\tfrac{1}{2}\left[\psi^-_{0}\left|\tfrac{1}{2},-\tfrac{1}{2}\right>\right]^{\sigma} \ , \qquad \qquad 
\tilde{\phi}_- = -\tfrac{1}{2}\left[ \psi^{-}_{-1} \chi^{-}_{0} \psi^{-}_{0} \left|\tfrac{1}{2},-\tfrac{1}{2}\right>\right]^{\sigma} = K^-_0 \tilde{\phi}_+ \ .
\end{equation}
In fact, we have $P_+ \tilde{\Psi}^j=\Psi^j$, where $P_+$ is the picture changing operator, see Appendix~\ref{appendix:superconformal-algebra-hybrid}.

\subsubsection{Free Bosons} \label{section:boson}

The four spacetime free bosons are singlets with respect to the spacetime $R$-sym\-metry. We will work in complex conventions, and for  the `barred' bosons we make the ansatz
\begin{equation} \label{eq:barred-bosons}
\partial \bar{\textbf{X}}^j:=\phi \, e^{i \sigma} \partial \bar{X}^j  \ ,
\end{equation}
in the ghost picture $P=0$. The physical conditions are
\begin{equation} \label{eq:boson-condition}
L_n \phi=0  \ \ (n \geq 0)\ , \qquad Q_{n} \phi=0 \ \ (n \geq -2) \ .
\end{equation}
Up to BRST exact states, the unique singlet state with spacetime conformal dimension $h=1$ is
\begin{equation} \label{eq:boson-state}
\phi = \left[ \chi^-_{-1} \chi^-_{0} \left| 0, \tfrac{1}{2}\right> \right]^{\sigma} \ .
\end{equation}
For the unbarred free bosons we make the ansatz
\begin{equation} \label{eq:unbarred-bosons}
\partial \textbf{X}^j:=\phi \, e^{4\rho +i \sigma + 2 i H} \partial X^j  \ .
\end{equation}
This ansatz has ghost picture $P=-4$. The physical state conditions are
\begin{equation} \label{eq:unbarred-condition}
L_n \phi=0  \ \ (n \geq 0)\ , \qquad Q_{n} \phi=0 \ \ (n \geq 2) \ .
\end{equation}
Again, up to BRST exact states, the unique singlet with spacetime conformal dimension $h=1$ is
\begin{equation} \label{eq:unbarred-state}
\phi =\left[ \psi^-_{-1} \psi^-_{0} \left| 1, -\tfrac{1}{2}\right> \right]^{\sigma} \ .
\end{equation}

\section{Correlators and Ward Identities}\label{sec:correlators}

In this section we use Ward identities to calculate the correlators of the worldsheet vertex operators that we have found in Section~\ref{section:states}; among other things this also shows that these vertex operators are not BRST trivial, something we have not analysed carefully above. We shall divide this section into two parts. First in Section~\ref{sec:exT}, we shall focus on the correlators that do not involve either of the two stress energy tensors of eqs.~(\ref{eq:t}) and (\ref{eq:sugawara-su2}); the correlators involving these fields are discussed in Section~\ref{sec:Tin}. The reason for this distinction is mainly technical: the worldsheet vertex operators associated to eqs.~(\ref{eq:t}) and (\ref{eq:sugawara-su2}) involve symplectic bosons with negative modes, while this is not the case for the other worldsheet states we have found above. For states without any negative symplectic boson modes the calculation of the correlators is significantly easier since we can basically just use the symplectic boson Ward identities of \cite{Dei:2020zui}; in the more general case of Section~\ref{sec:Tin}, on the other hand, we need to perform a more complicated analysis, similar to what was done in \cite{Bertle:2020sgd}.

As in \cite{Eberhardt:2019ywk,Dei:2020zui} the vertex operators on the worldsheet depend on an additional variable $x$, which is to be identified with the position in the dual CFT. More specifically, we define 
\begin{equation} \label{eq:vertex-operator}
V(\Phi,x,z)=e^{z L_{-1}} e^{x J^+_0} V(\Phi,0,0) e^{-x J^+_0} e^{-z L_{-1}} \ ,
\end{equation}
where $V(\Phi,0,0)$ is the vertex operator associated to $\Phi$ evaluated at $z=x=0$, and $L_{-1}$ is the world-sheet translation operator. We note that $[L_{-1},J^+_0]=0$, and therefore the order of $L_{-1}$ and $J^+_0$ in this definition is irrelevant. In the following sections, we will mainly focus on $\mathfrak{psu}(1,1|2)_1$ correlators and ignore the $z_i$-dependency that comes from the ghost sector. In order to soak up the $U_0$ charge we also need to introduce the vacuum $W$ field, see eq.~(\ref{eq:correlators-on-s2}) above. With respect to the symplectic boson fields it behaves as, see \cite[Section~2]{Dei:2020zui}
\begin{equation} \label{eq:eta-xi-w}
\xi^{\pm}(t) W(z) = \mathcal{O}(t-z) \ , \quad \quad \eta^{\pm}(t) W(z) = \frac{1}{(t-z)} V(\eta^{\pm}_{\frac{1}{2}} \left| 0 \right>^{(1)},z) + \mathcal{O}(1) \ .
\end{equation}

\subsection{Correlators excluding $\textbf{T}$ and $\textbf{T}_{\mathfrak{su}(2)}$}\label{sec:exT}
In Section~\ref{section:states} we found  the worldsheet vertex operators corresponding to the $\mathcal{N}=4$ superconformal generators, as well as to the free fermions and bosons on $\mathbb{T}^4$. Except for the stress energy tensors, all of these states are of the form 
\be\label{eq:genstate}
[F \left|m_1,m_2 \right>]^{\sigma} \ , 
\ee
where $F$ is a string of the fermionic modes $\chi^{\pm}$ and $\psi^{\pm}$. Since the fermions commute with the symplectic bosons, states of the form (\ref{eq:genstate}) satisfy the same symplectic boson Ward identities as in \cite{Dei:2020zui}. 

For the following it will also be convenient to use the Ward identities associated to the $\mathfrak{su}(2)_1 \subset \mathfrak{psu}(1,1|2)_1$ algebra. In particular, these constraints immediately imply that  $n$-point functions vanish unless the sum of $K^3_0$ eigenvalues is zero. Moreover, the $2$-point function of a singlet state with a state that sits in a doublet or a triplet representation of $\mathfrak{su}(2)$ must also be zero. This is already sufficient to show, for example, that 
\be
\left< \textbf{J}^a \textbf{T}\right>=0 \ , 
\ee
where $\textbf{J}^a$ are the states in Section~\ref{section:su2}, and $\textbf{T}$ is the stress energy tensor in eq.~(\ref{eq:t}). By the same token, it also follows that the $2$-point functions of $\textbf{T}$ with the supercurrents and fermions of Section~\ref{section:supercurrents} and Section~\ref{section:fermions} vanish. 

\subsubsection{The $\mathfrak{su}(2)_1$ generators}

We begin by considering the spacetime $R$-symmetry currents from Section~\ref{section:su2}. To start with we analyse their $2$-point functions. Using the $\mathfrak{su}(2)_1 \subset \mathfrak{psu}(1,1|2)_1$ Ward identities, the only non-zero correlators involving the vertex operators of $\mathfrak{su}(2)_1$ are $\left< \textbf{J}^3 \textbf{J}^3 \right>$ and $\left< \textbf{J}^{++} \textbf{J}^{--}\right>$, and they take the form\footnote{The $\mathfrak{psu}(1,1|2)_1$ conformal dimensions ($L_0$ eigenvalue) of the vertex operators in Section~\ref{section:su2} are zero. Therefore, these correlators are  $z_i$-independent.}
\begin{align}
\left< \textbf{J}^3(x_1,z_1) \textbf{J}^3(x_2,z_2) \right>&=\frac{d_{33}}{(x_1-x_2)^2} \ ,\\
\left< \textbf{J}^{++}(x_1,z_1) \textbf{J}^{--}(x_2,z_2) \right>&=\frac{d_{+-}}{(x_1-x_2)^2} \ .
\end{align}
Using $\mathfrak{su}(2)_1 \subset \mathfrak{psu}(1,1|2)_1$ Ward identities it furthermore follows that 
\begin{equation}
2d_{33}=d_{+-} \ .
\end{equation}
This matches with the spacetime theory, see Appendix~\ref{appendix:t4}.
\subsubsection{Fermions and Bosons}

Next we consider the correlation functions of the free fermions and bosons. For the calculation of the $2$-point functions of the fermions of Section~\ref{section:fermions}, it is convenient to consider $\tilde{\Psi}_{\pm}^j$ and $\Psi^j_{\pm}$, so that the sum of the ghost pictures is $P=-4$ as required by eq.~(\ref{eq:correlators-on-s2}), 
\begin{equation} \label{eq:fermions-correlators}
\left< \tilde{\phi}_{\pm} e^{2\rho+2i\sigma+i H^j} \phi_{\pm} e^{\rho+i\sigma+i H^k}\right> \ ,
\end{equation}
where $\phi_{\pm}$ and $\tilde{\phi}_{\pm}$ are defined in eqs.~(\ref{eq:fermions-+}) and (\ref{eq:fermions-p-+}), respectively. It is then immediate that the $2$-point functions involving two $\Psi_{\pm}^j$'s with the same value of $j$ are zero since they do not have the correct background charge on the sphere (see eq.~(3.9) of \cite{Dei:2020zui}). Thus the only non-zero correlators are $\langle \Psi_+^1 \Psi_-^2 \rangle$ and $\langle \Psi_+^2 \Psi_-^1 \rangle$. Using the explicit form of eq.~(\ref{eq:fermions-correlators}) we see\footnote{The minus sign comes from the Grassmann nature of the free fermions $e^{iH^j}$.}
\begin{equation} \label{fermions-minus-ahman}
\langle \Psi_+^1 (x_1,z_1) \Psi_-^2 (x_2,z_2) \rangle=-\langle \Psi_+^2(x_1,z_1) \Psi_-^1(x_2,z_2) \rangle \ .
\end{equation}
Using eq.~(\ref{eq:fermions-doublet-f}) this reproduces the expected correlators of the free fermions on $\mathbb{T}^4$. (Note that 
the minus sign in front of $\bar{\psi}^1$ in eq.~(\ref{eq:fermions-doublet-f}) comes from our convention for the $\mathfrak{su}(2)_1$ generators on $\mathbb{T}^4$, see Appendix~\ref{appendix:t4}.)

We can repeat the same calculation for the free bosons of Section~\ref{section:boson}. Again we note that the sum of the ghost pictures of the vertex operators in eqs.~(\ref{eq:barred-bosons}) and (\ref{eq:unbarred-bosons}) is $P=-4$ as required by eq.~(\ref{eq:correlators-on-s2}) for $2$-point functions. Now the only non-zero correlators are $\langle \partial \textbf{X}^j \partial \bar{\textbf{X}}^j \rangle$, and it follows directly from eq.~(\ref{eq:correlators-on-s2})  that they are indeed equal as expected.

\subsubsection{Supercurrents, Fermions and Bosons}

We close this section by discussing the $2$- and $3$-point functions of the supercurrents of Section~\ref{section:supercurrents}. For the $2$-point functions we consider the supercurrents $(G^+,\tilde{G}^-)$ as in eqs.~(\ref{eq:g+}) and (\ref{eq:gtilde-}), and the supercurrents $(\tilde{G}^+,G^-)$ as in eqs.~(\ref{eq:tildeG+ansatz1}) and (\ref{eq:g-}) so that the sum of the ghost pictures is $P=-4$, as required by eq.~(\ref{eq:correlators-on-s2}). Using $\mathfrak{su}(2)_1$ Ward identities we see that the only non-zero $2$-point functions of the supercurrents are $\left< G^+ G^- \right>$ and $\langle \tilde{G}^+ \tilde{G}^- \rangle$, and they take the form 
\begin{align} \label{s1}
\langle G^+(x_1,z_1) G^-(x_2,z_2) \rangle &\sim \frac{1}{(x_1-x_2)^3} \ , \\
\label{s2}
\langle \tilde{G}^+(x_1,z_1) \tilde{G}^-(x_2,z_2) \rangle&\sim \frac{1}{(x_1-x_2)^3} \ ,
\end{align}
where the $\sim$ indicates that these correlators depend now on $z_i$ --- this is a consequence of the fact that the $\mathfrak{psu}(1,1|2)_1$ conformal dimension ($L_0$ eigenvalue)  of the states in Section~\ref{section:supercurrents} are not zero. (Obviously, the full conformal dimension, including also the ghost contributions, is still zero.) With the conventions of Section~\ref{section:supercurrents} we find that 
\begin{equation}
\langle G^+(x_1,z_1) G^-(x_2,z_2) \rangle=\langle \tilde{G}^+(x_1,z_1) \tilde{G}^-(x_2,z_2) \rangle \ ,
\end{equation}
reproducing the expected structure of these correlators. 
\smallskip

Next we turn to the $3$-point functions of the supercurrents with bosons and fermions. The non-zero correlators are $\langle\partial\textbf{X}^i G^+ \Psi^j_{\pm} \rangle$ and $\langle \partial \textbf{X}^i \tilde{G}^- \Psi^j_{\pm}\rangle$ since the correlators of $G^+$ and $\tilde{G}^-$ with $\partial \bar{\textbf{X}}^i$ vanish. Using eq.~(\ref{eq:correlators-on-s2}) and setting $\Phi_1=\partial\textbf{X}^i$, $\Phi_2=G^+$ or $\Phi_2=\tilde{G}^{-}$, and $\Phi_3=\tilde{\Psi}_{\pm}^j$ we observe that the correlators are non-zero only if $i \neq j$ and the sum of $K^3_0$ eigenvalues is zero. We note that the sum of the ghost pictures is $P=-6$ as required by eq.~(\ref{eq:correlators-on-s2}) for $3$-point functions. For instance, if we set $i=1$, the only non-zero correlators are\footnote{In order not to clutter the notation, we shall in the following not write out the $(x_i,z_i)$ dependencies, and drop the $W$ field of eq.~(\ref{eq:correlators-on-s2}).}
\begin{equation}
\langle \partial \textbf{X}^1 G^+ \Psi^2_{-}\rangle \quad \quad \hbox{and} \quad \quad 
\langle \partial \textbf{X}^1 \tilde{G}^- \Psi^2_{+} \rangle \ .
\end{equation}
Using the $\mathfrak{su}(2)_1$ Ward identities we can relate them to one another, and with our conventions we find 
\begin{equation} \label{super-ferm-bose-1}
\langle \partial \textbf{X}^1 G^+ \Psi^2_{-}\rangle=\langle \partial \textbf{X}^1 \tilde{G}^- \Psi^2_{+}\rangle \ .
\end{equation}
Similarly for $i=2$ we can show
\begin{equation} \label{super-ferm-bose-3}
\langle \partial \textbf{X}^2 G^+ \Psi^1_{-}\rangle=\langle \partial \textbf{X}^2\tilde{G}^- \Psi^1_{+}\rangle \ .
\end{equation}
Moreover, because of a minus sign in $\tilde{G}^-_C$ between the $2$ bosons $\partial\bar{\textbf{X}}^k$ (see eq.~(\ref{eq:tildeGC})) we have
\begin{equation}
	\langle \partial \textbf{X}^1 G^+ \Psi^2_{-}\rangle=-\langle \partial \textbf{X}^2 \tilde{G}^- \Psi^1_{+}\rangle \ ,
\end{equation}
\begin{equation}
	\langle \partial \textbf{X}^2 G^+ \Psi^1_{-}\rangle=-\langle  \partial \textbf{X}^1 \tilde{G}^-\Psi^2_{+}\rangle  \ ,
\end{equation}
which matches with the spacetime theory, see Appendix~\ref{appendix:t4} and eq.~(\ref{eq:fermions-doublet-f}).

For the correlators involving $\tilde{G}^+$ and $G^-$ the situation is reversed in that the correlators with $\partial \textbf{X}^i$ vanish. It then follows from eq.~(\ref{eq:correlators-on-s2}) with $\Phi_1=\partial\bar{\textbf{X}}^i$, $\Phi_2=\tilde{G}^+$ or $\Phi_2=G^{-}$, and $\Phi_3=\Psi_{\pm}^j$, that the correlators 
are only non-zero for $i=j$ --- this follows from the requirement that the background charge is correct and the sum of the $K^3_0$ eigenvalues is zero. We again note the sum of the ghost pictures is $P=-6$, as required by eq.~(\ref{eq:correlators-on-s2}).
For example, if we set $i=1$, the only non-zero correlators are
\begin{equation} \label{eq:gtildepfermionboson}
\langle \partial \bar{\textbf{X}}^1 \tilde{G}^+ \Psi^1_{-}\rangle  \quad \quad \hbox{and} \quad \quad 
\langle \partial \bar{\textbf{X}}^1 G^- \Psi^1_{+}\rangle \ .
\end{equation}
Again, using the $\mathfrak{su}(2)_1$ Ward identities we can relate them to one another, and with our conventions  we get 
\begin{equation} \label{super-ferm-bose-2}
\langle \partial \bar{\textbf{X}}^1 \tilde{G}^+ \Psi^1_{-}\rangle =- 
\langle \partial \bar{\textbf{X}}^1 G^- \Psi^1_{+}\rangle \ .
\end{equation}
In a similar way, for $i=2$ we have
\begin{equation} \label{super-ferm-bose-4}
\langle \partial \bar{\textbf{X}}^2 \tilde{G}^+ \Psi^2_{-}\rangle =- 
\langle \partial \bar{\textbf{X}}^2 G^- \Psi^2_{+}\rangle \ .
\end{equation}
Again, because of a minus sign in $\tilde{G}^+_C$ between the $2$ bosons $\partial\textbf{X}^k$ (see eq.~(\ref{eq:tildeGC})) and the Grassmannian nature of the fermions $e^{i H^j}$, we get
\begin{equation}
\langle \partial \bar{\textbf{X}}^2 \tilde{G}^+ \Psi^2_{-}\rangle=-\langle \partial \bar{\textbf{X}}^1 G^- \Psi^1_{+}\rangle \ ,
\end{equation}
\begin{equation}
\langle \partial \bar{\textbf{X}}^1 \tilde{G}^+ \Psi^1_{-}\rangle=-\langle \partial \bar{\textbf{X}}^2 G^- \Psi^2_{+}\rangle \ .
\end{equation}
This reproduces what one expects from the spacetime $\mathbb{T}^4$ theory, see Appendix~\ref{appendix:t4} and eq.~(\ref{eq:fermions-doublet-f}).

\subsection{Correlators involving $\textbf{T}$ and $\textbf{T}_{\mathfrak{su}(2)}$}\label{sec:Tin}
\label{section:correlators-inc-t}

In the remainder of this section we analyse the correlation functions that also involve either of the two stress energy tensors, see eqs.~(\ref{eq:t}) and (\ref{eq:sugawara-su2}). As we mentioned before, their analysis is more complicated since the relevant states before spectral flow are not $\mathfrak{sl}(2,\mathds{R})$ highest weight states, and thus we cannot just use the Ward identities of \cite{Dei:2020zui}, but have to resort to the techniques of \cite{Bertle:2020sgd}. We shall consider four types of correlators. First, in Section~\ref{section:jjt}, we shall analyse the correlators involving $\textbf{T}$ together with $\textbf{J}^{++}$ and $\textbf{J}^{--}$, where $\textbf{J}^{\pm\pm}$ are the $\mathfrak{su}(2)_1$ generators of Section~\ref{section:su2}, and $\textbf{T}$ is the stress energy tensor in eq.~(\ref{eq:t}). This will, in particular, allow us to fix the coefficient $\alpha$ in eq.~(\ref{eq:t}). We shall then, in Section~\ref{section:t-twisted}, consider the correlators of $\textbf{T}$ with the $w$-twisted sector ground states $\Omega^w$ of the symmetric orbifold for small (odd) values of $w$. (The corresponding worldsheet states were already determined in \cite{Dei:2020zui}.) Finally, in Section~\ref{section:t-t}, we shall determine the $2$-point function of $\textbf{T}$ with itself. In Section~\ref{sec:tsu2}, we then turn to the correlators of the Sugawara $R$-symmetry stress energy tensor. In particular, we shall calculate the $2$-point functions of $\textbf{T}_{\mathfrak{su}(2)}$ with itself and $\textbf{T}$, $3$-point functions of $\textbf{T}_{\mathfrak{su}(2)}$ with $\mathfrak{su}(2)$ currents, fermions, and supercurrents, and in the end the $3$-point functions of $\textbf{T}_{\mathfrak{su}(2)}$ with the $w$-twisted ground states $\Omega^w$ for small (odd) values of $w$. Our results agree with the expectations from the symmetric orbifold, and hence support our identification of the above worldsheet states with these  spacetime fields. 

In this section, most of the vertex operators have the form (\ref{eq:comp-indep}), and thus the $3$-point functions of (\ref{eq:correlators-on-s2}) become, after removing the ghost contributions,
\begin{equation}
\left<  W \Phi_1 \Phi_2 (Q_{-1}\Phi_3) \right> \ .
\end{equation}

\subsubsection{$\left< W \textbf{J}^{++} \textbf{T} \textbf{J}^{--}\right>$} \label{section:jjt}

Let us begin by calculating the correlator
\begin{equation}
\left< W(u) \, \textbf{J}^{++}(x_1,z_1) \, \textbf{T}(x_2,z_2) \, (Q_{-1}\textbf{J}^{--})(x_3,z_3)\right> \ .
\end{equation}
In a first step we remove the negative symplectic boson modes from the different terms in $\textbf{T}$ by using contour integration techniques. Let us illustrate this  procedure for the third term in (\ref{eq:t}), which we write somewhat schematically as\footnote{In the following by $[\Phi]^{\sigma^w}$ or $[\Phi]^{\sigma^w}(x_i,z_i)$ we mean the vertex operator associated to this state at $(x_i,z_i)$. By slight abuse of notation we also mean by $\textbf{J}^{--}$ the vertex operator $\Phi^{--}$ without the ghost contribution, see eq.~(\ref{eq:comp-indep}), and similarly for the other vertex operators below.}
\begin{align} \label{eq:jpjmt}
\bigl< W & \textbf{J}^{++} \left[ \eta^+_{-2} \chi_0^- \psi_0^- \left| \tfrac{1}{2}, \tfrac{1}{2} \right> \right]^{\sigma} (Q_{-1}\textbf{J}^{--}) \bigr> \\ &= \int_{z_2} \frac{dt}{(t-z_2)^2} \left< \eta^{+}(t) W \textbf{J}^{++} \left[ \chi^-_0 \psi^-_0 \left| \tfrac{1}{2},\tfrac{1}{2} \right>\right]^{\sigma} (Q_{-1}\textbf{J}^{--}) \right> \notag \\
&= - \sum_{i\in \{1 ,3 \}} \int_{z_i} \frac{dt}{(t-z_2)^2} \left< \eta^{+}(t) W \textbf{J}^{++} \left[ \chi^-_0 \psi^-_0 \left| \tfrac{1}{2},\tfrac{1}{2} \right>\right]^{\sigma} (Q_{-1}\textbf{J}^{--})\right>\notag\\
&\quad - \int_{u} \frac{dt}{(t-z_2)^2} \left< \eta^{+}(t) W \textbf{J}^{++} \left[ \chi^-_0 \psi^-_0 \left| \tfrac{1}{2},\tfrac{1}{2} \right>\right]^{\sigma} (Q_{-1}\textbf{J}^{--}) \right>\notag \ .
\end{align}
Next we use the OPE of $\eta^+$ with $\textbf{J}^{++}$ 
\begin{equation}
\eta^+(z)\, \textbf{J}^{++}(x_1,z_1) = (z-z_1)^{-1} \Big[\eta^+_0 \left| \tfrac{1}{2},0 \right>\Big]^{\sigma}(x_1,z_1)  + {\cal O}(1) \ ,
\end{equation}
while the OPE of $\eta^+$ with $(Q_{-1} \textbf{J}^{--})$ involves only the action of $\eta^+_0$ on $(Q_{-1} \textbf{J}^{--})$ which again does not lead to any negative symplectic boson mode. The OPE of $\eta^+$ with $W$ is singular as in eq.~(\ref{eq:eta-xi-w})
\begin{equation} \label{eq:jpp-jmm-weta}
\frac{1}{z-u} \left< \Big[\eta^+_{\frac{1}{2}} W(u)\Big] \textbf{J}^{++} \left[ \chi^-_0 \psi^-_0 \left| \tfrac{1}{2},\tfrac{1}{2} \right>\right]^{\sigma} (Q_{-1}\textbf{J}^{--}) \right> \ .
\end{equation}
Combining these OPEs and using that none of the vertex operators that appear in the process have any negative symplectic boson modes, we can determine the correlators that appear on the right-hand-side of eq.~(\ref{eq:jpjmt}) using the methods of  \cite{Dei:2020zui}, see in particular Section~5.1.2 of that paper. Thus we can evaluate the contour integrals and hence calculate the left-hand-side of eq.~(\ref{eq:jpjmt}). The other terms that appear in $\textbf{T}$, see eq.~(\ref{eq:t}), can be dealt with similarly, and with the help of \texttt{Mathematica} and the Virasoro package \cite{mathematica} we obtain altogether
\begin{equation}
\frac{\left< W(u) \textbf{J}^{++}(x_1,z_1) \textbf{T}(x_2,z_2) (Q_{-1}\textbf{J}^{--})(x_3,z_3)\right>}{\left< W(u) \textbf{J}^{++}(x_1,z_1) \Omega(x_2,z_2) (Q_{-1}\textbf{J}^{--})(x_3,z_3)\right>}=\frac{2 \alpha (x_3-x_1)^2}{(x_1-x_2)^2 (x_2-x_3)^2} \ .
\end{equation}
Thus the correct normalisation of the stress energy tensor is 
\begin{equation} \label{eq:normalization-of-t}
\alpha = \frac{1}{2} \ .
\end{equation}
We have checked that this normalisation then also leads to the correct answer for the correlators where we replace $\textbf{J}^{\pm\pm}$ by the free fermions, the free bosons or the supercurrents.

\subsubsection{$\left< W \Omega^w \textbf{T} \Omega^w\right>$}
\label{section:t-twisted}

Now that we have fixed the normalisation of $\textbf{T}$ we can test our ansatz further by calculating the $3$-point functions with the ground states of the $w$-twisted sectors $\Omega^w$. In particular, this should reproduce the correct conformal dimensions of the twisted sector ground states; the following is the hybrid formalism version of the calculation that was performed in \cite{Bertle:2020sgd}. 

More concretely, let us consider the correlators 
\begin{equation}
\left< W(u) \, \Omega^w(x_1,z_1) \, \textbf{T}(x_2,z_2) \, (Q_{-1}\Omega^w)(x_3,z_3)\right> \ ,
\end{equation}
where we take $w>1$ to be odd.\footnote{The analysis for even $w$ is a bit more complicated.} The state that corresponds to the $w$-twisted ground state is (for $w$ odd) \cite{Dei:2020zui}
\begin{equation} \label{eq:w-twistedgroundstate}
\Omega^w = \left[ \chi^-_{-\frac{w-1}{2}} \psi^-_{-\frac{w-1}{2}} \cdots \chi^-_{0} \psi^-_{0} \left| m_1,m_2 \right> \right]^{\sigma^w} e^{2\rho+i\sigma+i H} \ ,
\end{equation}
where 
\begin{equation}
m_1+m_2 = - \frac{w^2+1}{4w} \ , \qquad \qquad m_1 - m_2 = \frac{1}{2} \ .
\end{equation}
As before, see eq.~(\ref{eq:jpjmt}), we remove the negative symplectic boson modes from $\textbf{T}$ using contour integration techniques, and then evaluate the resulting correlators further --- once the negative symplectic boson modes have been removed, this can be done as in \cite{Dei:2020zui}. Some of the details of this calculation are spelled out in Appendix~\ref{app:E.1}, and for all the $w$ we have tested, i.e.\ $w \in \{ 1, 3, 5, 7 , 9 \}$, we find (with $\alpha=\frac{1}{2}$)
\begin{equation}
\frac{\left< W(u) \Omega^w(x_1,z_1) \textbf{T}(x_2,z_2) (Q_{-1}\Omega^w)(x_3,z_3)\right>}{\left< W(u) \Omega^w(x_1,z_1) \Omega(x_2,z_2) (Q_{-1}\Omega^w)(x_3,z_3)\right>}= \frac{w^2-1}{4 w^2}\frac{(x_3-x_1)^2}{(x_1-x_2)^2 (x_2-x_3)^2} \ .
\end{equation}
As in eq.~(2.17) of \cite{Bertle:2020sgd}, this therefore leads to the correct conformal dimension associated to the $w$-twisted ground state.

\subsubsection{$\left< \textbf{T} \textbf{T} \right>$}
\label{section:t-t}

We have also calculated the $2$-point function of the stress energy tensor with itself. The method that we use for calculating $\left< \textbf{T}(x_1,z_1) \textbf{T}(x_2,z_2) \right>$ is the same as in the previous subsection, and the more specific details are explained in Appendix~\ref{app:E.2}. We find 
\begin{equation} \label{eq:2-t-t}
\frac{\left< \textbf{T}(x_1,z_1) \textbf{T}(x_2,z_2) \right>}{\left< \Omega(x_1,z_1) \Omega(x_2,z_2) \right>}=\frac{12 \alpha^2}{(x_1-x_2)^4}  = \frac{3}{(x_1-x_2)^4}=\frac{c/2}{(x_1-x_2)^4} \ ,
\end{equation}
where we have first set $\alpha=\frac{1}{2}$ from eq.~(\ref{eq:normalization-of-t}). Thus, again, our worldsheet calculation reproduces the correct spacetime answer since the central charge of the torus theory is $c=6$.

\subsubsection{$\textbf{T}_{\mathfrak{su}(2)}$} \label{sec:tsu2}
As a final test we study the correlators involving the worldsheet vertex operator associated to the Sugawara stress energy tensor of the spacetime $R$-symmetry, namely
\be
[\Phi_{T_{\mathfrak{su(2)}}}]^{\sigma} e^{2\rho+i\sigma+iH} \ ,
\ee
where $\Phi_{T_{\mathfrak{su(2)}}}$ is given in eq.~(\ref{eq:sugawara-su2}). The state in the first line of eq.~(\ref{eq:sugawara-su2}) is just the Sugawara stress energy tensor associated to the $\mathfrak{su}(2)_1$ subalgebra of $\mathfrak{psu}(1,1|2)_1$ (provided we set $\beta=\tfrac{1}{4}$, as we shall do from now on)
\be \label{eq:sugawara-tsu2}
\Phi^F_{\mathfrak{su}(2)}=\tfrac{1}{3} \big[  K^3_{-1} K^3_{-1}+\tfrac{1}{2} (K^+_{-1} K^-_{-1} + K^-_{-1} K^+_{-1})\big] \Omega_{1,\frac{1}{2}} \ .
\ee
The remaining states in eq.~(\ref{eq:sugawara-su2}) equal
\begin{align} \label{eq:sugawara-null}
\Phi^S_{\mathfrak{su}(2)}=\tfrac{1}{2}[ \, &\xi^-_{-2} \chi_0^- \psi_0^- \left| \tfrac{3}{2}, \tfrac{1}{2} \right>
+\, \xi^+_{-2} \chi_0^- \psi_0^- \left| 1, 0 \right>
+\, \xi^-_{-1} \xi^-_{-1} \chi_0^- \psi_0^- \left| 2, \tfrac{1}{2} \right> \\
&+ \, \xi^+_{-1} \xi^-_{-1} \chi_0^- \psi_0^- \left| \tfrac{3}{2}, 0 \right> \, ]^{\sigma} \ . \notag
\end{align}
It is not difficult to see that $\Phi^S_{\mathfrak{su}(2)}$ is an $\mathfrak{su}(2)_1$ highest weight state with spin zero, and thus, its  $2$-point function with $\Phi^F_{\mathfrak{su}(2)}$ vanishes. We have also checked that the $2$-point function of $\Phi^S_{\mathfrak{su}(2)}$ with itself is zero, again using techniques similar to those of Appendix~\ref{app:E.2}. As a consequence, the $2$-point function of $\textbf{T}_{\mathfrak{su}(2)}$ with itself equals that of $\Phi^F_{\mathfrak{su}(2)}$, and thus we find
\begin{equation}\label{eq:4.38}
\frac{\left<\textbf{T}_{\mathfrak{su}(2)}(x_1,z_1) \textbf{T}_{\mathfrak{su}(2)}(x_2,z_2)\right>}{\left<\Omega(x_1,z_1) \Omega(x_2,z_2)\right>} = \frac{1/2}{(x_1-x_2)^4} = \frac{c_{\mathfrak{su}(2)}/2}{(x_1-x_2)^4} \ ,
\end{equation}
where we have used that the central charge of the Sugawara stress energy tensor of $\mathfrak{su}(2)_1$ is $c_{\mathfrak{su}(2)}=1$.

Next we turn to calculating the $2$-point function of $\textbf{T}$ with $\textbf{T}_{\mathfrak{su}(2)}$. 
With respect to the worldsheet $\mathfrak{su}(2)_1$ algebra, the vertex operator associated to $\textbf{T}$ is a highest weight state with spin zero, and thus the $\mathfrak{su}(2)_1$ Ward identities imply that $\langle \textbf{T} \, \Phi^F_{\mathfrak{su}(2)} \rangle=0$. On the other hand, the $2$-point function of $ \textbf{T}$ with $\Phi^S_{\mathfrak{su}(2)}$ is non-zero, and it can be calculated using the same techniques as in Section~\ref{section:t-t}. This then leads to  
\begin{equation}\label{eq:4.39}
\frac{\left<\textbf{T}(x_1,z_1) \textbf{T}_{\mathfrak{su}(2)}(x_2,z_2)\right>}{\left<\Omega(x_1,z_1) \Omega(x_2,z_2)\right>} = \frac{1/2}{(x_1-x_2)^4} \ ,
\end{equation}
which is  the expected answer from the spacetime perspective, see Appendix~\ref{appendix:t4}. (From that viewpoint, it follows from the usual coset argument that $\langle (\textbf{T} - \textbf{T}_{\mathfrak{su}(2)})\, \textbf{T}_{\mathfrak{su}(2)} \rangle=0$, and hence eq.~(\ref{eq:4.39}) must agree with (\ref{eq:4.38}).)

We have also analysed the $3$-point functions of $\textbf{T}_{\mathfrak{su}(2)}$ with the bosons, $\mathfrak{su}(2)$ currents, fermions and supercurrents. In all cases we have found that the contribution from $\Phi^S_{\mathfrak{su}(2)}$ vanishes, and thus the correlators can be simply evaluated using the $\mathfrak{su}(2)_1$ Ward identities. It is then for example immediate that the $3$-point functions of $\textbf{T}_{\mathfrak{su}(2)}$ with the bosons from Section~\ref{section:boson} vanish because they are singlets with respect to $\mathfrak{su}(2)_1 \subset \mathfrak{psu}(1,1|2)_1$; this is also what one expects from the spacetime perspective. Similarly, we can work out the $3$-point functions of $\textbf{T}_{\mathfrak{su}(2)}$ with the spacetime $R$-symmetry currents, fermions and supercurrents of Section \ref{section:states}, and we find\footnote{The $Q_n$ modes in eqs.~(\ref{eq:sugawara-correlators}) and (\ref{eq:tsu2-w}) only modify the behavior of the vertex operators with respect to the global $\mathfrak{su}(2)$ subalgebra of $\mathfrak{psu}(1,1|2)_1$.}
\begin{subequations} \label{eq:sugawara-correlators}
\begin{align}
\frac{\langle W(u) \textbf{J}^{++}(x_1,z_1) \textbf{T}_{\mathfrak{su}(2)}(x_2,z_2)(Q_{-1}\textbf{J}^{--})(x_3,z_3)\rangle}{\langle W(u) \textbf{J}^{++}(x_1,z_1) \Omega(x_2,z_2) (Q_{-1}\textbf{J}^{--})(x_3,z_3)\rangle}&=\frac{(x_3-x_1)^2}{(x_1-x_2)^2 (x_2-x_3)^2} \ ,
\\
\frac{\langle W(u) \Psi_+^2(x_1,z_1) \textbf{T}_{\mathfrak{su}(2)}(x_2,z_2)(Q_0 \tilde{\Psi}_-^1)(x_3,z_3)\rangle}{\langle W(u) \Psi_+^2(x_1,z_1) \Omega(x_2,z_2)(Q_0 \tilde{\Psi}_-^1)(x_3,z_3)\rangle}&=\frac{1}{4}\frac{(x_3-x_1)^2}{(x_1-x_2)^2 (x_2-x_3)^2} \ ,
\\
\frac{\langle W(u) \Psi_+^1(x_1,z_1) \textbf{T}_{\mathfrak{su}(2)}(x_2,z_2)(Q_0 \tilde{\Psi}_-^2)(x_3,z_3)\rangle}{\langle W(u) \Psi_+^1(x_1,z_1) \Omega(x_2,z_2)(Q_0 \tilde{\Psi}_-^2)(x_3,z_3)\rangle}&=\frac{1}{4}\frac{(x_3-x_1)^2}{(x_1-x_2)^2 (x_2-x_3)^2} \ ,
\\
\frac{\langle W(u) G^+(x_1,z_1) \textbf{T}_{\mathfrak{su}(2)}(x_2,z_2)(Q_2 G^-)(x_3,z_3)\rangle}{\langle W(u) G^+(x_1,z_1) \Omega(x_2,z_2)(Q_2 G^-)(x_3,z_3)\rangle}&=\frac{1}{4}\frac{(x_3-x_1)^2}{(x_1-x_2)^2 (x_2-x_3)^2} \ ,
\\
\frac{\langle W(u) \tilde{G}^-(x_1,z_1) \textbf{T}_{\mathfrak{su}(2)}(x_2,z_2)(Q_2\tilde{G}^+)(x_3,z_3)\rangle}{\langle W(u) \tilde{G}^-(x_1,z_1) \Omega(x_2,z_2)(Q_2\tilde{G}^+)(x_3,z_3)\rangle}&=\frac{1}{4}\frac{(x_3-x_1)^2}{(x_1-x_2)^2 (x_2-x_3)^2} \ .
\end{align}
\end{subequations}
This then also correctly reproduces the $3$-point functions of the spacetime theory.

Finally, we have calculated the $3$-point functions of $\textbf{T}_{\mathfrak{su}(2)}$ with $w$-twisted ground states for odd $w$. Following the analysis of eq.~(2.15) in \cite{Bertle:2020sgd}, the spacetime theory predicts that the $3$-point functions for $w$ odd should equal
\begin{equation} \label{eq:tsu2-w}
\frac{\langle W(u) \Omega^w(x_1,z_1) \textbf{T}_{\mathfrak{su}(2)}(x_2,z_2) (Q_{-1}\Omega^w)(x_3,z_3)\rangle}{\langle W(u) \Omega^w(x_1,z_1) \Omega(x_2,z_2) (Q_{-1}\Omega^w)(x_3,z_3)\rangle}= \frac{w^2-1}{24 w^2}\frac{(x_3-x_1)^2}{(x_1-x_2)^2 (x_2-x_3)^2} \ ,
\end{equation}
where we have again used that the central charge of the Sugawara stress energy tensor is $c_{\mathfrak{su}(2)}=1$. In order to calculate the $3$-point functions in eq.~(\ref{eq:tsu2-w}) from the worldsheet, we first note that $\Omega^w$ in eq.~(\ref{eq:w-twistedgroundstate}) for odd $w$ is a singlet with respect to the $\mathfrak{su}(2)_1$ subalgebra of $\mathfrak{psu}(1,1|2)_1$. Therefore, the contribution from $\Phi^F_{\mathfrak{su}(2)}$ is zero for odd $w$. For the calculation of the $3$-point functions of $\Phi^S_{\mathfrak{su}(2)}$ with $\Omega^w$ we use the same techniques as in Section~\ref{section:t-twisted}, see also Appendix~\ref{app:E.1}. Our final result for $w \in \{1,3,5,7,9\}$ matches with the result from the spacetime theory in eq.~(\ref{eq:tsu2-w}).

\section{Conclusions}\label{sec:conclusions}

In this paper we have considered string theory on ${\rm AdS}_3 \times {\rm S}^3 \times \mathbb{T}^4$ with one unit ($k=1$) of NS-NS flux. We have worked with the hybrid formalism of \cite{Berkovits:1999im} and utilised that the $\mathfrak{psu}(1,1|2)_1$ factor at level $k=1$ has a free field realisation. We have then analysed systematically the low-lying physical states. In particular, we have managed to identify the worldsheet vertex operators corresponding to the ${\cal N}=4$ superconformal fields, as well as the free fermions and free bosons of the dual CFT (the symmetric orbifold of $\mathbb{T}^4$). We have confirmed these identifications by calculating correlators involving these fields, thereby showing that they reproduce what is expected from the dual CFT. 

One of the nice structural results that we have found is that the $R$-symmetry of the dual ${\cal N}=4$ superconformal 2d CFT can indeed be identified with the global $\mathfrak{su}(2)$ subalgebra of $\mathfrak{psu}(1,1|2)$ on the worldsheet. (In particular, we have checked that all the fields we have constructed respect this identification.) While this identification is what one should have expected on general grounds, it is far from obvious from a technical viewpoint: on the face of it, the $R$-symmetry of the dual symmetric orbifold comes from the bilinears in the fermions of the $\mathbb{T}^4$, and therefore does not seem to have anything to do with the $\mathfrak{su}(2)$ subalgebra of $\mathfrak{psu}(1,1|2)$ (under which the worldsheet fermions of the $\mathbb{T}^4$ factor  are invariant). Somehow the BRST cohomology seems to `glue' together these two  $\mathfrak{su}(2)$ symmetries, and it would be very interesting to understand more conceptually how this identification comes about. 

The BRST analysis of \cite{Berkovits:1999im} is quite complicated, but we have noticed that the physical spectrum has a significantly simpler form if one does not perform the final similarity transformation of \cite{Berkovits:1999im}, see eq.~(4.14) of that paper.  We suspect that this may also be useful for future attempts to determine the physical spectrum of this hybrid theory more completely. In particular, it would be nice to prove directly that the physical spectrum reproduces that of the symmetric orbifold; the argument that was given in \cite{Eberhardt:2018ouy} was rather indirect. In any case, given that a natural generalisation of our free field hybrid model seems to be the worldsheet theory describing free ${\cal N}=4$ SYM in 4d \cite{Gaberdiel:2021iil,Gaberdiel:2021jrv}, it will be important to develop powerful techniques to study worldsheet theories of this type.

\section*{Acknowledgments}

We thank Rajesh Gopakumar and Vit Sriprachyakul for useful discussions, and Andrea Dei and Bob Knighton for comments on a draft version of the paper. This paper is based on the Master thesis of one of us (K.N.) \cite{Master}. This research was supported by a personal grant from the Swiss National Science Foundation, as well as through the NCCR SwissMAP that is also funded by the Swiss National Science Foundation.

\appendix

\section{Free field realisation of $\mathfrak{psu}(1,1|2)_1$}
\label{appendix:free-field}

Let us review the free field realisation of $\mathfrak{psu}(1,1|2)_1$  in terms of symplectic bosons and fermions  following \cite{Eberhardt:2018ouy,Dei:2020zui}. We have four symplectic bosons $(\xi^{\pm},\eta^{\pm})$ and four (real) fermions $(\chi^{\pm},\psi^{\pm})$ with OPEs 
\begin{equation} \label{symplectic-bosons-commutation}
\xi^{\alpha}(z) \eta^{\beta}(w) \sim \frac{\epsilon^{\alpha \beta}}{(z-w)} \ , \qquad 
\psi^{\alpha}(z) \chi^{\beta}(w)\sim \frac{\epsilon^{\alpha \beta}}{(z-w)} \ , 
\end{equation}
where $\alpha,\beta=\pm$, $\epsilon^{+-}=-\epsilon^{-+}=1$, and $\epsilon^{++}=\epsilon^{--}=0$. The $\mathfrak{u}(1,1|2)_1$ currents can then be written as 
\begin{align} \label{eq:definition-psu-com}
J^3(z)& =-\tfrac{1}{2}(\eta^+ \xi^-)(z)-\tfrac{1}{2}(\eta^- \xi^+)(z)  \quad & J^{\pm}(z)=(\eta^{\pm} \xi^{\pm})(z) \\
K^3(z) & =-\tfrac{1}{2}(\chi^+ \psi^-)(z)-\tfrac{1}{2}(\chi^- \psi^+)(z) \quad &  K^{\pm}(z)=\pm(\chi^{\pm} \psi^{\pm})(z) \label{su2-psu-minusdare}  \\
S^{\alpha\beta+}(z) & =(\xi^{\alpha} \chi^{\beta})(z) \quad &  S^{\alpha\beta-}(z)=-(\eta^{\alpha}\psi^{\beta})(z) \ ,
\end{align}
and their modes satisfy the commutation relations ($k=1$)
\begin{subequations} \label{psu-affine}
\begin{align}
[J^3_n,J^3_m] = & -\tfrac{1}{2} k n \delta_{n+m,0}\\
[J^3_n,J^{\pm}_m] = & \pm J^{\pm}_{n+m}\\
[J^+_n,J^-_m]= & k n \delta_{n+m,0}-2 J^3_{n+m}\\
[K^3_n,K^3_m] = & \tfrac{1}{2} k n \delta_{n+m,0}\\
[K^3_n,K^{\pm}_m] = & \pm K^{\pm}_{n+m}\\
[K^+_n,K^-_m]= & k n \delta_{n+m,0}+2 K^3_{n+m}\\
[J^a_n,S^{\alpha\beta\gamma}_m]= & \tfrac{1}{2} c_a (\sigma^a)^{\alpha}_{\mu} S^{\mu \beta\gamma}_{n+m}\label{psu-supercurrents-sl2}\\
[K^a_n,S^{\alpha\beta\gamma}_m]= &\tfrac{1}{2} (\sigma^a)^{\beta}_{\nu} S^{\alpha \nu \gamma}_{n+m} \label{psu-supercurrents-multiplet}\\
\{S^{\alpha \beta\gamma}_n,S^{\mu \nu\rho}_m\}= & k n \epsilon^{\alpha \mu} \epsilon^{\beta \nu} \epsilon^{\gamma \rho} \delta_{n+m,0}-\epsilon^{\beta \nu} \epsilon^{\gamma \rho} c_a \sigma_a^{\alpha \mu} J^a_{n+m}\\&+ \epsilon^{\alpha \mu} \epsilon^{\gamma \rho} \sigma_a^{\beta \nu} K^a_{n+m} + \epsilon^{\alpha \mu} \epsilon^{\beta \nu} \delta^{\gamma,-\rho} Z_{n+m} \label{psu-supercurrents} \ ,
\end{align}
where $Z=U+V$ and $Y=U-V$, and the $\mathfrak{u}(1)$ currents $U$ and $V$ are defined via 
\be
U(z)=-\tfrac{1}{2}(\eta^+ \xi^-)(z)+\tfrac{1}{2}(\eta^- \xi^+)(z) \ , \quad  V(z)=-\tfrac{1}{2}(\chi^+ \psi^-)(z)+\tfrac{1}{2}(\chi^- \psi^+)(z) \ .
\ee
\end{subequations}
Here the $\sigma$ matrices are explicitly 
\begin{equation} \label{psu-sigma-def-1}
(\sigma^-)^+_-=2\ , \quad (\sigma^3)^-_-=-1\ , \quad (\sigma^3)^+_+=1\ , \quad (\sigma^+)^-_+=2 \ , 
\end{equation}
\begin{equation} \label{psu-sigma-def-2}
(\sigma_-)^{--}=1\ , \quad (\sigma_3)^{-+}=1\ , \quad (\sigma_3)^{+-}=1\ , \quad (\sigma_+)^{++}=-1 \ . 
\end{equation}
Furthermore, $c_a=-1$ for $a=-$ and $c_a=1$ for $a=+,3$. Note that with respect to the $U$ and $V$ currents, the free fields carry the charges 
\begin{subequations} \label{uv}
\begin{gather}
[U_n,\xi^{\pm}_{m}]=-\tfrac{1}{2} \xi^{\pm}_{n+m} \ , \qquad  [U_n,\eta^{\pm}_{m}]=\tfrac{1}{2} \eta^{\pm}_{n+m} \ , \\
[V_n,\psi^{\pm}_{m}]=-\tfrac{1}{2} \psi^{\pm}_{n+m} \ , \qquad [V_n,\chi^{\pm}_{m}]=\tfrac{1}{2} \chi^{\pm}_{n+m} \ .
\end{gather}
\end{subequations}
In order to go from $\mathfrak{u}(1,1|2)$ to $\mathfrak{psu}(1,1|2)$, we have to set $Z_n=0$. Technically this is achieved by working with the subspace of states on which $Z_n$ with $n\geq 0$ vanishes, $Z_n\phi=0$; the $Z_{-n}$ descendants are then null, and thus automatically quotiented out. 

The representations that appear in the worldsheet theory are the Ramond sector representation (where all modes are integer moded), as well as its spectrally flowed images. On the Ramond ground states we define the action of the symplectic boson zero modes via 
\begin{subequations} \label{psu-symplectic-rep}
\begin{align}
\xi^+_0 \left| m_1,m_2 \right> = \left| m_1 ,m_2+\tfrac{1}{2} \right\rangle \ , \qquad & \eta^+_0 \left| m_1,m_2 \right> =2 m_1 \left| m_1 +\tfrac{1}{2},m_2 \right\rangle \ ,\\
\xi^-_0 \left| m_1,m_2 \right> = -\left| m_1-\tfrac{1}{2} ,m_2 \right\rangle \ , \qquad &  \eta^-_0 \left| m_1,m_2 \right> =-2 m_2 \left| m_1,m_2-\tfrac{1}{2} \right\rangle \ . 
\end{align}
\end{subequations}
As regards the fermionic zero modes we require 
\be
\psi^+_0\, \left| m_1,m_2 \right> =  \chi^+_0  \, \left| m_1,m_2 \right> = 0 \ . 
\ee
Then 
\be
K^3_0 \, \left| m_1,m_2 \right> = \tfrac{1}{2} \, \left| m_1,m_2 \right> \ . 
\ee
We also use the convention, see \cite{Dei:2020zui}, that 
\be
U_0\, \left| m_1,m_2 \right> = (m_1 - m_2 - \tfrac{1}{2})\, \left| m_1,m_2 \right> \ , \qquad 
V_0\, \left| m_1,m_2 \right> = 0 \ .
\ee
\smallskip

\noindent The  $\mathfrak{psu}(1,1|2)_k$ algebra has a spectral flow automorphism 
\begin{subequations} \label{psu-spectral-flow}
\begin{align}
\sigma^w(J^3_n)& =J^3_n+\tfrac{kw}{2}\delta_{n,0} \label{psu-j3-sigma}\\
\sigma^w(J^{\pm}_n)& =J^{\pm}_{n\mp w} \label{psu-jpm-sigma}\\
\sigma^w(K^3_n) & =K^3_n+\tfrac{kw}{2}\delta_{n,0}\label{psu-sigma-w}\\
\sigma^w(K^{\pm}_n)& =K^{\pm}_{n\pm w} \\
\sigma^w(S^{\alpha\beta\gamma}_n)& =S^{\alpha\beta\gamma}_{n+\frac{w}{2}(\beta-\alpha)} \ , 
\end{align}
\end{subequations}
which is induced, for $k=1$, by the spectral flow action 
\be
\sigma(\xi^\pm_n) = \xi^\pm_{n \mp \frac{1}{2}} \ , \quad 
\sigma(\eta^\pm_n) = \eta^\pm_{n \mp \frac{1}{2}} \ , \quad 
\sigma(\psi^\pm_n) = \psi^\pm_{n \pm \frac{1}{2}} \ , \quad 
\sigma(\chi^\pm_n) = \chi^\pm_{n \pm \frac{1}{2}} 
\ee
on the underlying free fields. On the worldsheet Virasoro algebra spectral flow acts as 
\begin{equation} \label{spectral-psu-ln}
\sigma^w(L_n)=L_n+w(K^3_n-J^3_n) \ .
\end{equation}
This allows us to define spectrally flowed representations of $\mathfrak{psu}(1,1|2)_1$: let $\phi$ be any state in a (highest weight) representation of $\mathfrak{psu}(1,1|2)_1$, then the states of the $w$-fold spectrally flowed representation are denoted by $[\phi]^{\sigma^w}$.  They form a representation of $\mathfrak{psu}(1,1|2)_1$ where we define the action of $S_n \in \mathfrak{psu}(1,1|2)_1$ via
\begin{equation} \label{eq:spectral-flow-representation}
S_n [\phi]^{\sigma^w} =[\sigma^w(S_n) \phi]^{\sigma^w} \ .
\end{equation}
Thus the spectrally flowed representation has `the same' underlying vector space, but the action of the $\mathfrak{psu}(1,1|2)_1$ modes has been `twisted' by $\sigma^w$.

\section{The torus theory}
\label{appendix:t4}

Let us start with the RNS formulation of the $\mathbb{T}^4$ theory. It is convenient to work with the complex fields 
\begin{equation} \label{n=4-scalar-ope}
\partial X^j (z) \partial \bar{X}^k (w) = \frac{\delta^{jk}}{(z-w)^2} \ , \qquad 
\psi^j (z) \bar{\psi}^k (w) = \frac{\delta^{jk}}{(z-w)} \ ,
\end{equation}
where $j,k \in \{1,2\}$. The theory has a natural $\mathcal{N}=2$ superconformal symmetry generated by the fields 
\begin{eqnarray} \label{n=4-t}
T  = \partial X^j \partial \bar{X}^k - \tfrac{1}{2}\bar{\psi}^j \partial \psi^j - \tfrac{1}{2} \psi^j \partial \bar{\psi}^j \ , \qquad  & J = \tfrac{1}{2} \psi^j \bar{\psi}^j \ , \\ 
G^{+}=\partial \bar{X}^j \psi^j \ , \qquad & 
G^{-}=\partial X^j \bar{\psi}^j \ . \label{g+-t4}
\end{eqnarray}
They satisfy the OPEs of an $\mathcal{N}=2$ superconformal algebra
\begin{subequations} \label{eq:n=2-convention}
	\begin{align}
	&J(z) G^{\pm}(w) = \pm\frac{1}{2}\frac{G^{\pm}(w)}{(z-w)} \ , \quad \quad \quad J(z) J(w) = \frac{c/12}{(z-w)^2} \ ,
	\\&G^+(z) G^-(w) = \frac{c/3}{(z-w)^3}+\frac{2 J(w)}{(z-w)^2}+\frac{T(w)+\partial J(w)}{(z-w)} \ , \label{eq:n=2-g+g-}
	\end{align}
\end{subequations}
where $c$ is the central charge (in our case $c=6$). The topological twist amounts to modifying the stress-energy tensor to 
\be\label{toptwist}
T_C = T +  \partial J \ .
\ee
With respect to $T_C$, the supercurrents $G^+$ and $G^-$ then have conformal dimension $1$ and $2$, respectively. (After the topological twist we shall denote the corresponding fields by $G^\pm_C$; in terms of the free fields they are still given by the above expressions, but their conformal dimensions have now been shifted. We shall use this convention also for the other fields of the ${\cal N}=2$ and ${\cal N}=4$ algebra, i.e.\ they will carry an index $C$ if we think of them as being part of the topologically twisted algebra.)

The ${\cal N}=2$ algebra is actually part of a (small)  $\mathcal{N}=4$ superconformal algebra; in particular, we can enhance the $\mathfrak{u}(1)$ algebra generated by $J$ to an $\mathfrak{su}(2)$ algebra whose additional currents are 
\begin{equation} \label{j++--t4}
J^{++} = \psi^1 \psi^2 \ , \qquad 
J^{--} = - \bar{\psi}^1 \bar{\psi}^2 \ .
\end{equation}
The OPE of these currents with the original supercurrents generate another set of supercurrents 
\begin{equation} \label{eq:tildeGC}
\tilde{G}^{+}_C = - \epsilon_{ij} \partial X^i \psi^j \ , \qquad 
\tilde{G}^{-}_C = - \epsilon_{ij} \partial \bar{X}^i \bar{\psi}^j \ .
\end{equation}
Altogether these fields then generate an ${\cal N}=4$ superconformal symmetry. 
In the hybrid formulation one bosonises the fermions via 
\begin{equation}
	\psi^j=e^{i H^j}\ , \qquad  \bar{\psi}^j=e^{-i H^j} \ , 
\end{equation}
where $j=1,2$. 
We also define 
\be\label{Hdef}
H = H^1 + H^2 \ .
\ee
In terms of these free fields the $\mathfrak{su}(2)$ currents of the topologically twisted theory are then 
\be\label{JC}
J_C = \tfrac{1}{2} i \partial H \  , \qquad J^{++}_C=e^{i H} \ , \qquad 
J^{--}_C=-e^{-i H} \ . 
\ee

\section{The ${\cal N}=4$ superconformal algebra}
\label{appendix:superconformal-algebra-hybrid}

The hybrid worldsheet theory possesses an $\mathcal{N}=2$ superconformal algebra with generating fields \cite{Berkovits:1999im}\footnote{We thank Vit Sriprachyakul for helping us check the conventions in eqs.~(\ref{eq:hybrid-n=2-beforesim}).}
\begin{subequations} \label{eq:hybrid-n=2-beforesim}
\begin{align}
T& =T_{\mathfrak{psu}(1,1|2)_1}-\tfrac{1}{2} \left( (\partial \rho)^2 + (\partial \sigma)^2 \right) + \tfrac{3}{2} \partial^2 (\rho + i \sigma) + T_C \\ 
G^{+} & =e^{-\rho} Q + e^{i \sigma} T + G^{+}_C -\partial (e^{i\sigma} J_{\rho H}) \label{Gp} \\
G^{-} & =e^{-i \sigma} \label{Gm} \\ 
J & = \tfrac{1}{2} J_{\rho \sigma} + J_C \ ,
\end{align}
\end{subequations}
where $J_{\rho\sigma}=\partial(\rho+i\sigma)$, $J_{\rho H}= \partial \rho + 2 J_C$, and the fields with index $C$ arise from the topologically twisted torus theory, see Appendix~\ref{appendix:t4} for details. In terms of the free-field realisation of $\mathfrak{psu}(1,1|2)_1$ of Appendix~\ref{appendix:free-field} we have \cite{Dei:2020zui}
\begin{equation} \label{eq:q}
Q=2 (\chi^+ \chi^-) (\xi^+ \partial \xi^- - \xi^- \partial \xi^+) \ . 
\end{equation}
We should stress that these formulae correspond to eq.~(4.13) of \cite{Berkovits:1999im}, i.e.\ we have \emph{not} performed the similarity transformation of eq.~(4.14) of \cite{Berkovits:1999im}. For the BRST analysis this is obviously irrelevant, but we have found that the physical states take a simpler form with this choice of BRST operator. 

The fields in eqs.~(\ref{eq:hybrid-n=2-beforesim}) satisfy a twisted $\mathcal{N}=2$ algebra. This is to say, they have the same OPE relations as in eqs.~(\ref{eq:n=2-convention}), except for the
OPE of eq.~(\ref{eq:n=2-g+g-}) that is replaced by 
\begin{equation}
	G^+(z) G^-(w) = \frac{c/3}{(z-w)^3}+\frac{2 J(w)}{(z-w)^2}+\frac{T(w)}{(z-w)} \ .\label{eq:n=2-twisted}
\end{equation}
Furthermore, because of the twist of eq.~(\ref{toptwist}), $J$ is now not primary w.r.t.\ $T$ any more, 
\begin{equation}
	T(z) J(w) = \frac{-c/6}{(z-w)^3}+\frac{J(w)}{(z-w)^2}+\frac{\partial J(w)}{(z-w)} \ ,
\end{equation}
where $c=6$ is the central charge. In order to extend this ${\cal N}=2$ algebra to ${\cal N}=4$ we introduce the  $\mathfrak{su}(2)$ currents 
\begin{equation}
J^{++} = e^{\rho + i \sigma} J^{++}_C \ , \qquad 
J^{--} = - e^{-\rho - i \sigma} J^{--}_C \ , 
\end{equation}
where $J^{++}_C$ and $J^{--}_C$ were defined in (\ref{JC}). The second set of supercurrents are then \cite{Berkovits:1999im}
\begin{align} \label{tilde-g+-hybrid}
\tilde{G}^{+} & =e^{\rho+i H} \\ 
\tilde{G}^{-} & = e^{-2\rho-i \sigma - i H} Q - e^{-\rho - i H} T - e^{-\rho-i\sigma} \tilde{G}^{-}_C +e^{-\rho-iH} (i \partial \sigma J_{\rho H}+\partial J_{\rho H})\ . \label{tilde-g--hybrid}
\end{align}
We define the ghost picture as the eigenvalue of the operator
\begin{equation} \label{eq:ghost-picture} 
P = \frac{1}{2 \pi i} \int dz\, \bigl [(\eta \xi)(z)-(\beta \gamma)(z) \bigr] = \frac{1}{2 \pi i} \int dz\, \bigl [-\partial \phi + i \partial \kappa \bigr] \ , 
\end{equation}
where $\beta\gamma$ are the commuting ghosts of the RNS string (with conformal dimensions $3/2$ and $-1/2$, respectively), while $\xi\eta$ is the $bc$ system that appears in the bosonisation of $\beta\gamma$, see Appendix~\ref{app:boso}, and we have used the bosonisation formulae from there in the final step. A vertex operator of the form $\phi e^{m\rho+in \sigma + i k_1 H^1 +i k_2 H^2}$ therefore has ghost number 
\begin{equation} \label{ghost-calc}
\Phi = \phi \, e^{m\rho+in \sigma + i k_1 H^1 +i k_2 H^2} : \qquad 
P(\Phi) =2(k_1+k_2)-3m \ ,
\end{equation}
where we have used that $\rho$ is defined via \cite[eq.~(4.11)]{Berkovits:1999im}
\begin{equation} \label{rhodef}
\rho = -2\phi - i \kappa - i (H^{1,\text{RNS}} +H^{2,\text{RNS}}) \ , \quad \hbox{and} \quad i H^i= i H^{i,\text{RNS}}+\phi+i\kappa \ . 
\end{equation}
The picture changing operator $P_+$ maps physical states to physical states and is explicitly given as, see \cite[eq.~(13.71)]{Lust:1989tj}
\begin{equation} \label{eq:picture-changing}
P_+ = \frac{-1}{2 \pi i} \int dz \, G^+\,  (\xi \cdot \circ) \ ,
\end{equation}
where we mean by $\xi \cdot \circ$ that we first take the normal ordered product  with $\xi$ before taking the integral. Using the bosonisation of $\xi$ from Appendix~\ref{app:boso} this becomes 
\begin{equation} \label{picture-changing-hybrid}
P_+ = \frac{-1}{2 \pi i} \int dz \, G^+ \, e^{-\rho - i H}   \ , 
\end{equation}
where we have used the bosonisation formula (\ref{xibos}), (\ref{rhodef}), as well as (\ref{Hdef}).

\section{$bc$ systems and bosonisation}\label{app:bc}
In this appendix, we review some basic facts about $bc$ systems and their bosonisation, see e.g.\ \cite{Lust:1989tj} for a good introduction. 

\label{appendix:bosonization}
\subsection{$bc$ systems}
A $bc$ system consists of two primary fields $b$ and $c$ with conformal dimensions $\lambda$ and $1-\lambda$, respectively. Their OPEs are
\begin{equation} \label{bc-ope}
b(z) c(w) = \frac{\epsilon}{(z-w)} \ , \quad \quad c(z) b(w) = \frac{1}{(z-w)} \ ,
\end{equation}
where $\epsilon=1$ if both fields are fermionic and $\epsilon=-1$ if both are bosonic. The $bc$ ghost number current $J_{bc}=(bc)(z)$ has the property 
\begin{equation}
J_{bc}(z) b(w) = \frac{b(w)}{(z-w)} \ , \qquad  J_{bc}(z) c(w) = \frac{-c(w)}{(z-w)} \ , 
\end{equation}
and therefore the ghost number of $b$ is $+1$, while that of $c$ is $-1$. More generally, we define the ghost number operator by the zero mode of $J_{bc}$, 
\begin{equation}
P_{bc}=\frac{1}{2\pi i} \int dz J_{bc}(z)  \ . 
\end{equation}
The energy-momentum tensor of the $bc$ system is 
\begin{equation} \label{bc-t}
T_{bc}(z)=(\partial b)c\, (z)-\lambda\,  \partial(bc)(z) \ , 
\end{equation}
so that $b$ and $c$ are primary with conformal dimension $\lambda$ and $1-\lambda$, respectively. The central charge of the system can be read off from the OPE of 
$T_{bc}$ with itself, and one finds 
\begin{equation} \label{bc-c}
c_{bc}=-2 \epsilon \, \bigl (6 \lambda (\lambda-1)+1 \bigr) \ . 
\end{equation}
We also note that the ghost number current $J_{bc}$ is not primary with respect to $T_{bc}$ unless $\lambda=\frac{1}{2}$, since one calculates
\begin{equation} \label{bc-t-j}
T_{bc}(z) J_{bc}(w)=\frac{\epsilon(2\lambda-1)}{(z-w)^3}+\frac{J(w)}{(z-w)^2}+\frac{\partial J(w)}{(z-w)} \ . 
\end{equation}

\subsection{Bosonisation of $bc$ systems}\label{app:boso}

In the description of the RNS string two $bc$ systems appear: an anti-commuting ($\epsilon=1$) system with $\lambda=2$, corresponding to the conformal symmetry; and a commuting ($\epsilon=-1)$ system with $\lambda=\frac{3}{2}$, reflecting the supercurrent constraint. Following standard conventions, we shall denote the first system by $bc$, and the  second by $\beta\gamma$. Let us describe their bosonisations in turn. 
The $bc$ system can be bosonised by writing, see e.g.\ \cite{Lust:1989tj} 
\begin{equation}
	b(z)=e^{-i\sigma(z)}\ , \qquad  c(z)=e^{i \sigma(z)} \ ,
\end{equation}
where $\sigma$ is a boson satisfying
\begin{equation}
\sigma(z)\sigma(w)=-\text{ln}(z-w) \ . 
\end{equation}
The energy-momentum tensor of this bosonised theory is
\begin{equation}
T^{\sigma}=-\frac{1}{2} (\partial \sigma)^2+\frac{3}{2}\partial^2 (i\sigma) \ ,
\end{equation}
and it is straightforward to check that the conformal dimension of $b$ and $c$ are $2$ and $-1$, respectively. After bosonisation the ghost number current takes the simple form
\begin{equation}
J_{bc}(z)=-i \partial \sigma(z) \ . 
\end{equation}
For the commuting $\beta\gamma$ system bosonisation is a bit more complicated: we write 
\begin{equation}
\beta(z)=((\partial \xi) e^{-\phi})(z) \ , \quad \quad \gamma(z)=(\eta e^{\phi})(z) \ ,
\end{equation}
where $\phi$ is a boson with the usual OPE
\begin{equation}
\phi(z)\phi(w)=-\text{ln}(z-w) \ ,
\end{equation}
and $\xi \eta$ is an anti-commuting $bc$ system with conformal dimensions $1$ and $0$ respectively ($\epsilon=+1$, $\lambda=1$).\footnote{The $\xi$ and $\eta$ fields are not to be confused with the symplectic boson fields $\xi^\pm$ and $\eta^\pm$ that appear in the free field realisation of $\mathfrak{psu}(1,1|2)_1$.} To complete the bosonisation, we then also bosonise the $\xi \eta$ system as 
\begin{equation}\label{xibos}
\xi(z)=e^{-i \kappa(z)}\ , \qquad  \eta(z)=e^{i \kappa(z)} \ , 
\end{equation}
where $\kappa$ is yet another boson with OPE 
\begin{equation}
\kappa(z)\kappa(w)=-\text{ln}(z-w) \ .
\end{equation}
The ghost number current of the $\beta\gamma$ system is then 
\be
J_{\beta\gamma}(z) = \beta\gamma  (z) = \partial \phi (z) \ .
\ee

\section{Ward Identities in the Correlators of $\textbf{T}$}
\label{appendix:t}
In this appendix, we explain our method for calculating the correlators that involve the stress energy tensor, see Section~\ref{section:correlators-inc-t}.

\subsection{The correlator with the $w$-twisted ground state}\label{app:E.1}
Let us begin by analysing the correlator of the stress energy tensor with $w$-twisted ground states, see Section~\ref{section:t-twisted},
\begin{equation} \label{eq:t-w-w}
\left< W(u) \Omega^w(x_1,z_1) \textbf{T}(x_2,z_2) (Q_{-1}\Omega^w)(x_3,z_3)\right> \ .
\end{equation}
The worldsheet vertex operator associated to the stress energy tensor is given in eq.~(\ref{eq:t}), and the corresponding state involves negative symplectic boson modes. They can be dealt with as described in Section~\ref{section:correlators-inc-t}. Let us illustrate this for the term $[\xi^+_{-2} \chi_0^- \psi_0^- \left| 1, 0 \right>]^{\sigma}$ in the stress energy tensor (\ref{eq:t}), for which we write\footnote{Here we have omitted the $Q_{-1}$ mode acting on $\Omega^w$ in eq.~(\ref{eq:t-w-w}) since it does not modify the symplectic boson Ward identities of \cite{Dei:2020zui}.}
\begin{align} \label{omega-w-example}
\!\!\left< W \Omega^w \left[ \xi^+_{-2} \chi^-_0 \psi^-_0 \left| 1,0 \right>\right]^{\sigma} \Omega^w \right> &= \int_{z_2} \frac{dt}{(t-z_2)^2} \left< \xi^{+}(t) W \Omega^w \left[ \chi^-_0 \psi^-_0 \left| 1,0 \right>\right]^{\sigma} \Omega^w \right> \\
&= - \sum_{i\in \{1 ,3 \}} \int_{z_i} \frac{dt}{(t-z_2)^2} \left< \xi^{+}(t) W \Omega^w \left[ \chi^-_0 \psi^-_0 \left| 1,0 \right>\right]^{\sigma} \Omega^w \right> .\notag  
\end{align}
Next we use the OPE  of $\xi^+$ with the $w$-twisted ground state
\begin{equation} \label{ope-xi-w}
\xi^+ (t) \Omega^w (x_i,z_i)= \sum_{j=\frac{1}{2}}^{\frac{w}{2}} (t-z_i)^{-j-\frac{1}{2}} (\xi^+_{j} \Omega^w)(x_i,z_i) \ ,
\end{equation}
which involves the `unknowns' 
\begin{subequations} \label{eq:unknowns}
\begin{align}
\left< W(u) \Omega^w(x_1,z_1) V_{n_1,n_2}(x_2,z_2) (\xi^+_l \Omega^w)(x_3,z_3)\right>  \ &, \\
\left< W(u) (\xi^+_l \Omega^w)(x_1,z_1) V_{n_1,n_2}(x_2,z_2) \Omega^w(x_3,z_3)\right> \ & ,
\end{align}
\end{subequations}
where $l=\frac{1}{2},\ldots \frac{w-1}{2}$, and $V_{m_1,m_2}$ is the vertex operator associated to $\left[\left|m_1,m_2\right>\right]^{\sigma}$. Note that $V_{m_1,m_2}$ satisfies the same symplectic boson Ward identities \cite{Dei:2020zui} as $\Omega_{m_1,m_2}$, see eq.~(\ref{eq:omegam1m2}). In order to determine the `unknowns' in eqs.~(\ref{eq:unknowns}) we use the Ward identities coming from the insertion of $\xi^\pm(z)$. Because of Liouville's theorem we have 
\begin{align}
&\left< \xi^+(z) W(u) \Omega^w(x_1,z_1) V_{n_1,n_2}(x_2,z_2) \Omega^w(x_3,z_3)\right>\\ &\quad =\sum_{l=\frac{1}{2}}^{\frac{w}{2}} (z-z_1)^{-l-\frac{1}{2}} \left< W(u) (\xi^+_l \Omega^w)(x_1,z_1) V_{n_1,n_2}(x_2,z_2) \Omega^w(x_3,z_3)\right> \notag\\
&\qquad +\sum_{l=\frac{1}{2}}^{\frac{w}{2}} (z-z_3)^{-l-\frac{1}{2}} \left< W(u) \Omega^w(x_1,z_1) V_{n_1,n_2}(x_2,z_2) (\xi^+_l \Omega^w)(x_3,z_3)\right>\notag\\
&\qquad +(z-z_2)^{-1} \left< W(u) \Omega^w(x_1,z_1) V_{n_1,n_2+\frac{1}{2}}(x_2,z_2) \Omega^w(x_3,z_3)\right>\notag \ ,
\end{align}
and similarly for $\xi^-(z)$
\begin{align}
&\left< \xi^-(z) W(u) \Omega^w(x_1,z_1) V_{n_1,n_2}(x_2,z_2) \Omega^w(x_3,z_3)\right>\\&\quad = -\sum_{l=\frac{1}{2}}^{\frac{w}{2}} \frac{x_1}{(z-z_1)^{l+\frac{1}{2}}} \left< W(u) (\xi^+_l \Omega^w)(x_1,z_1) V_{n_1,n_2}(x_2,z_2) \Omega^w(x_3,z_3)\right> \notag\\
&\qquad -\sum_{l=\frac{1}{2}}^{\frac{w}{2}} \frac{x_3}{(z-z_3)^{l+\frac{1}{2}}} \left< W(u) \Omega^w(x_1,z_1) V_{n_1,n_2}(x_2,z_2) (\xi^+_l \Omega^w)(x_3,z_3)\right>\notag\\
&\qquad -\frac{x_2}{(z-z_2)} \left< W(u) \Omega^w(x_1,z_1) V_{n_1,n_2+\frac{1}{2}}(x_2,z_2) \Omega^w(x_3,z_3)\right>\notag \ .
\end{align}
As in \cite{Eberhardt:2019ywk,Dei:2020zui} we now observe that the OPE of $(\xi^-(z) + x_i \xi^+(z))$ with $\Omega^w$ is of order $(z-z_i)^{\frac{w-1}{2}}$ for $i \in \{ 1,3 \}$, while it is of order $\mathcal{O}(1)$ for $z\sim z_2$, and of order $(z-u)$ for $z\sim u$. Moreover, we know the coefficient of the term of order $(z-z_i) ^{\frac{w-1}{2}}$ for $i \in \{ 1,3 \}$ at $z=z_i$ in terms of the zero mode action of $\xi^-$ before spectral flow \cite{Dei:2020zui}. Thus we have enough constraints to determine the `unknown' terms, and hence can determine the corelator in (\ref{omega-w-example}).\footnote{Recently, the general solution to these recursion relations was found in \cite{Dei:2021xgh}.}

The other terms can be dealt with similarly, except that the analysis is a bit more complicated in general. In particular, for the term of the form $ \xi^+_{-1} \eta^+_{-1} \chi_0^- \psi_0^- \left| \tfrac{1}{2}, 0 \right>$, we need to repeat the above argument twice, and then end up with `unknowns' of the form 
\begin{subequations} \label{unk-xi-eta}
\begin{align} \label{unk-xi-eta-21}
\langle W(u) (\eta^+_j \xi^+_l \Omega^w)(x_1,z_1) V_{n_1,n_2}(x_2,z_2) \Omega^w(x_3,z_3)\rangle \ &, \\ \label{unk-xi-eta-22}
\langle W(u) (\xi^+_l \Omega^w)(x_1,z_1) V_{n_1,n_2}(x_2,z_2) (\eta^+_j \Omega^w)(x_3,z_3)\rangle \ &, \\ \label{unk-xi-eta-23}
\langle (\eta^{\pm}_{\frac{1}{2}} W)(u) (\xi^+_{j} \Omega^w)(x_1,z_1) V_{n_1,n_2}(x_2,z_2) \Omega^w(x_3,z_3)\rangle \ &, \\ \label{unk-xi-eta-24}
\langle W(u) \Omega^w(x_1,z_1) V_{n_1,n_2}(x_2,z_2) (\eta^+_j \xi^+_l\Omega^w)(x_3,z_3)\rangle \ &, \\ \label{unk-xi-eta-25}
\langle W(u) (\eta^{+}_j \Omega^w)(x_1,z_1) V_{n_1,n_2}(x_2,z_2) (\xi^+_l \Omega^w)(x_3,z_3)\rangle \ &, \\ \label{unk-xi-eta-26}
\langle (\eta^{\pm}_{\frac{1}{2}} W)(u) \Omega^w(x_1,z_1) V_{n_1,n_2}(x_2,z_2) (\xi^+_{j}\Omega^w)(x_3,z_3)\rangle \ &.
\end{align}
\end{subequations}
In order to calculate the unknowns (\ref{unk-xi-eta-21}), (\ref{unk-xi-eta-22}) and (\ref{unk-xi-eta-23}) we proceed as in \cite{Bertle:2020sgd}. This is to say, we fix $j$ and consider the correlators
\begin{align}
&\left< \eta^+(z) W(u) (\xi^+_{j} \Omega^w)(x_1,z_1) V_{n_1,n_2}(x_2,z_2) \Omega^w(x_3,z_3)\right>\\
&\quad =(z-u)^{-1} \left< (\eta^+_{\frac{1}{2}} W)(u) (\xi^+_{j} \Omega^w)(x_1,z_1) V_{n_1,n_2}(x_2,z_2) \Omega^w(x_3,z_3)\right>\notag\\
&\qquad +\sum_{l=\frac{1}{2}}^{\frac{w}{2}} (z-z_1)^{-l-\frac{1}{2}} \left< W(u) (\eta^+_l \xi^+_{j} \Omega^w)(x_1,z_1) V_{n_1,n_2}(x_2,z_2) \Omega^w(x_3,z_3)\right> \notag\\
&\qquad +\sum_{l=\frac{1}{2}}^{\frac{w}{2}} (z-z_3)^{-l-\frac{1}{2}} \left< W(u) (\xi^+_{j} \Omega^w)(x_1,z_1) V_{n_1,n_2}(x_2,z_2) (\eta^+_l \Omega^w)(x_3,z_3)\right>\notag\\
&\qquad +2 n_1 (z-z_2)^{-1} \left< W(u) (\xi^+_{j} \Omega^w)(x_1,z_1) V_{n_1+\frac{1}{2},n_2}(x_2,z_2) \Omega^w(x_3,z_3)\right>\notag \ ,
\end{align}
and 
\begin{align}
&\left< \eta^-(z) W(u) (\xi^+_{j} \Omega^w)(x_1,z_1) V_{n_1,n_2}(x_2,z_2) \Omega^w(x_3,z_3)\right>\\
&\quad =(z-u)^{-1} \left< (\eta^-_{\frac{1}{2}} W)(u) (\xi^+_{j} \Omega^w)(x_1,z_1) V_{n_1,n_2}(x_2,z_2) \Omega^w(x_3,z_3)\right>\notag\\
&\qquad -\sum_{l=\frac{1}{2}}^{\frac{w}{2}} \frac{x_1}{(z-z_1)^{l+\frac{1}{2}}} \left< W(u) (\eta^+_l \xi^+_{j} \Omega^w)(x_1,z_1) V_{n_1,n_2}(x_2,z_2) \Omega^w(x_3,z_3)\right>\notag \\
&\qquad -\sum_{l=\frac{1}{2}}^{\frac{w}{2}} \frac{x_3}{(z-z_3)^{l+\frac{1}{2}}} \left< W(u) (\xi^+_{j} \Omega^w)(x_1,z_1) V_{n_1,n_2}(x_2,z_2) (\eta^+_l \Omega^w)(x_3,z_3)\right>\notag\\
&\qquad -2 n_1 \frac{x_2}{(z-z_2)} \left< W(u) (\xi^+_{j} \Omega^w)(x_1,z_1) V_{n_1+\frac{1}{2},n_2}(x_2,z_2) \Omega^w(x_3,z_3)\right>\notag \ .
\end{align}
Then we can use the fact that the OPE of $(\eta^-(z) + x_i \eta^+(z))$ is of the order $(z-z_i)^{\frac{w-1}{2}}$ for $i=3$, while it is of order $\mathcal{O}(1)$ for $z\sim z_2$. For $z\sim z_1$ the situation is now a bit more complicated since the term of order $(z-z_1)^{j-\frac{1}{2}}$ is not zero, but rather equals
\begin{equation}
-(z-z_1)^{j-\frac{1}{2}}\left< W(u) \Omega^w(x_1,z_1) V_{n_1,n_2}(x_2,z_2) \Omega^w(x_3,z_3)\right> \ .
\end{equation}
In addition, we know the coefficient of the leading order term for all $i \in \{1,2,3\}$ since it is given by the action of $\eta^-_0$ before spectral flow. Thus we have again enough equations to determine the unknowns of this type. The same method can also be applied to the unknowns (\ref{unk-xi-eta-24}), (\ref{unk-xi-eta-25}) and (\ref{unk-xi-eta-26}), and thus we can also solve these more complicated correlators. This allows us to determine the full correlator (\ref{eq:t-w-w}).

\subsection{The $2$-point function of $\textbf{T}$}\label{app:E.2}
In this appendix, we explain our method for calculating the $2$-point function of the stress energy tensor, see Section~\ref{section:t-t},
\begin{equation}
	\left< \textbf{T}(x_1,z_1) \textbf{T}(x_2,z_2) \right> \ .
\end{equation}
As in Appendix~\ref{app:E.1}, see eq.~(\ref{omega-w-example}), we first use the usual contour deformation argument to rewrite the negative mode of the symplectic boson in terms of a contour integral of the corresponding symplectic boson field. For example, for one of the terms we get 
\begin{align} \label{example-t-t}
&\left< \left[ \xi^+_{-2} \chi_0^- \psi_0^- \left| 1, 0 \right> \right]^{\sigma} (x_1,z_1) \, \left[ \eta^-_{-2} \chi_0^- \psi_0^- \left| 1, 1 \right> \right]^{\sigma} (x_2,z_2) \right>\\
&=-\int_{z_2} \frac{dt}{(t-z_2)^2} \left< \xi^+(t) \left[ \chi_0^- \psi_0^- \left| 1, 0 \right> \right]^{\sigma} (x_1,z_1)\, \left[ \eta^-_{-2} \chi_0^- \psi_0^- \left| 1, 1 \right> \right]^{\sigma} (x_2,z_2) \right>\notag \ .
\end{align}
Then we use the OPE 
\begin{align}
&\xi^+ (t) \left[ \eta^-_{-2} \chi_0^- \psi_0^- \left| 1, 1 \right> \right]^{\sigma}(x_2,z_2) = \sum_{r} (t-z_2)^{-r-1} \left[ \xi^+_{r} \eta^-_{-2} \chi_0^- \psi_0^- \left| 1, 1 \right> \right]^{\sigma}\\
&=(t-z_2)^{-1} \left[ \eta^-_{-2} \chi_0^- \psi_0^- \left| 1, \tfrac{3}{2} \right> \right]^{\sigma}+(t-z_2)^{-3} \left[ \chi_0^- \psi_0^- \left| 1, 1 \right> \right]^{\sigma} + {\cal O}(t-z_2)\notag \ .
\end{align}
We can repeat the same procedure for $\eta^{-}_{-2}$ and thereby express the 
correlator in eq.~(\ref{example-t-t}) as a sum of terms (with different values for $m_i$ and $n_i$)
\begin{equation} \label{eq:2-chi-psi}
\left< \left[ \chi_0^- \psi_0^- \left| m_1, m_2 \right> \right]^{\sigma}(x_1,z_1) \left[ \chi_0^- \psi_0^- \left| n_1, n_2 \right> \right]^{\sigma}(x_2,z_2) \right> \ .
\end{equation}
These correlators now satisfy the same symplectic boson Ward identities as the correlators of highest weight states
\begin{equation} \label{main-t-t}
\left< V_{m_1, m_2}(x_1,z_1) V_{n_1, n_2}(x_2,z_2) \right> \ .
\end{equation}
More specifically, we have following \cite{Dei:2020zui}
\be \label{xip-laurent}
\langle \xi^{+}(t) V_{m_1, m_2}(x_1,z_1) V_{n_1, n_2}(x_2,z_2) \rangle=\frac{\langle [\xi^+_{1/2} V_{m_1, m_2}] V_{n_1, n_2} \rangle}{(t-z_1)}+\frac{\langle V_{m_1, m_2}[ \xi^+_{1/2} V_{n_1, n_2}] \rangle}{(t-z_2)} \ ,
\ee
and
\be\label{xim-laurent}
\langle \xi^{-}(t) V_{m_1, m_2}(x_1,z_1) V_{n_1, n_2}(x_2,z_2) \rangle =-x_1 \frac{\langle [\xi^+_{1/2} V_{m_1, m_2}] V_{n_1, n_2} \rangle}{(t-z_1)}-x_2\frac{\langle V_{m_1, m_2}[ \xi^+_{1/2} V_{n_1, n_2}] \rangle}{(t-z_2)}  \ .
\ee
These correlators satisfy
\be  \label{xim-1}
\bigl\langle \bigl(\xi^{-}(t)+x_1 \xi^+(t)\bigr) V_{m_1, m_2}(x_1,z_1) V_{n_1, n_2}(x_2,z_2) \bigr\rangle =\langle [\xi^-_{-1/2} V_{m_1, m_2}] V_{n_1, n_2} \rangle  \quad \text{as} \quad t \rightarrow z_1  \ ,
\ee
and
\be\label{xim-2}
\bigl\langle \bigl(\xi^{-}(t)+x_2 \xi^+(t)\bigr) V_{m_1, m_2}(x_1,z_1) V_{n_1, n_2}(x_2,z_2) \bigr\rangle =\langle V_{m_1, m_2}[ \xi^-_{-1/2} V_{n_1, n_2}] \rangle  \quad \text{as} \quad t \rightarrow z_2 \ .
\ee
Imposing the conditions (\ref{xim-1}) and (\ref{xim-2}) on the correlators in eqs.~(\ref{xip-laurent}) and (\ref{xim-laurent}) leads to 
\begin{subequations} \label{eq:xi2point}
\be
\frac{x_2-x_1}{z_2-z_1} \langle V_{m_1, m_2+\frac{1}{2}}(x_1,z_1) V_{n_1, n_2}(x_2,z_2) \rangle +\langle V_{m_1, m_2}(x_1,z_1) V_{n_1-\frac{1}{2}, n_2}(x_2,z_2) \rangle=0  \ ,
\ee
and
\be
\frac{x_2-x_1}{z_2-z_1} \langle V_{m_1, m_2}(x_1,z_1) V_{n_1, n_2+\frac{1}{2}}(x_2,z_2) \rangle+\langle V_{m_1-\frac{1}{2}, m_2}(x_1,z_1) V_{n_1, n_2}(x_2,z_2) \rangle=0  \ .
\ee
\end{subequations}
We can do the same for $\eta^{\pm}$, and thereby obtain the equations 
\begin{subequations} \label{eq:eta2point}
\be
m_1 \frac{x_2-x_1}{z_2-z_1} \langle V_{m_1+\frac{1}{2}, m_2}(x_1,z_1) V_{n_1, n_2}(x_2,z_2) \rangle+n_2 \langle V_{m_1, m_2}(x_1,z_1) V_{n_1, n_2-\frac{1}{2}}(x_2,z_2) \rangle=0  \ ,
\ee
and
\be
n_1 \frac{x_2-x_1}{z_2-z_1} \langle V_{m_1, m_2}(x_1,z_1) V_{n_1+\frac{1}{2}, n_2}(x_2,z_2) \rangle +m_2 \langle V_{m_1, m_2-\frac{1}{2}}(x_1,z_1) V_{n_1, n_2}(x_2,z_2) \rangle=0  \ .
\ee
\end{subequations}
This fixes the correlators up to an overall normalisation constant 
\begin{equation} \label{2point-formula}
\langle V_{m_1, m_2}(x_1,z_1) V_{n_1, n_2}(x_2,z_2) \rangle=a\, \frac{(-1)^{2 m_1+2 n_1} \, \delta_{m_2-n_1,-\frac{1}{2}} \delta_{m_1-n_2,\frac{1}{2}} }{(x_1-x_2)^{h_1+h_2} (z_1-z_2)^{\Delta_1+\Delta_2}} \ ,
\end{equation}
where $h_i$ is the spacetime conformal dimension (the eigenvalue of $J^3_0$) associated to the corresponding state, i.e.\ $h_1 = m_1 + m_2 + \frac{1}{2}$, and $h_2 = n_1 + n_2 + \frac{1}{2}$, 
and $\Delta_i$ is the worldsheet conformal dimension (the eigenvalue of $L_0$); here we have fixed the $z$-dependence by the usual worldsheet Ward identities. 

\makeatletter
\g@addto@macro\bfseries{\boldmath}
\makeatother

\advance\textheight by 1.0cm


\begin{thebibliography}{99}

\bibitem{Eberhardt:2018ouy}
L.~Eberhardt, M.R.~Gaberdiel and R.~Gopakumar,
``The Worldsheet Dual of the Symmetric Product CFT,''
JHEP {\bf 1904} (2019) 103
{\tt  [\href{https://arxiv.org/abs/1812.01007}{arXiv:1812.01007 [hep-th]}]}. 

\bibitem{Giveon:2005mi}
A.~Giveon, D.~Kutasov, E.~Rabinovici and A.~Sever,
``Phases of quantum gravity in AdS$_3$ and linear dilaton backgrounds,''
Nucl.\ Phys.\ B \textbf{719} (2005) 3
{\tt  [\href{https://arxiv.org/abs/hep-th/0503121}{hep-th/0503121}]}.

\bibitem{Gaberdiel:2017oqg}
M.R.~Gaberdiel, R.~Gopakumar and C.~Hull,
``Stringy AdS$_{3}$ from the worldsheet,''
JHEP \textbf{1707} (2017) 090
{\tt  [\href{https://arxiv.org/abs/1704.08665}{arXiv:1704.08665 [hep-th]}]}. 

\bibitem{Ferreira:2017pgt}
K.~Ferreira, M.R.~Gaberdiel and J.I.~Jottar,
``Higher spins on AdS$_{3}$ from the worldsheet,''
JHEP \textbf{1707} (2017) 131
{\tt  [\href{https://arxiv.org/abs/1704.08667}{arXiv:1704.08667 [hep-th]}]}. 

\bibitem{Giribet:2018ada}
G.~Giribet, C.~Hull, M.~Kleban, M.~Porrati and E.~Rabinovici,
``Superstrings on AdS$_{3}$ at $k=1$,''
JHEP {\bf 1808} (2018) 204
{\tt  [\href{https://arxiv.org/abs/1803.04420}{arXiv:1803.04420 [hep-th]}]}. 

\bibitem{Gaberdiel:2018rqv}
M.R.~Gaberdiel and R.~Gopakumar,
``Tensionless string spectra on AdS$_{3}$,''
JHEP {\bf 1805} (2018) 085
{\tt  [\href{https://arxiv.org/abs/1803.04423}{arXiv:1803.04423 [hep-th]}]}. 

\bibitem{Eberhardt:2019ywk}
L.~Eberhardt, M.R.~Gaberdiel and R.~Gopakumar,
``Deriving the AdS$_{3}$/CFT$_{2}$ correspondence,''
JHEP \textbf{2002} (2020) 136
{\tt [\href{https://arxiv.org/abs/1911.00378}{arXiv:1911.00378 [hep-th]}]}.

\bibitem{Eberhardt:2020akk}
L.~Eberhardt,
``AdS$_{3}$/CFT$_{2}$ at higher genus,''
JHEP \textbf{2005} (2020) 150
{\tt [\href{https://arxiv.org/abs/2002.11729}{arXiv:2002.11729 [hep-th]}]}.

\bibitem{Dei:2020zui}
A.~Dei, M.R.~Gaberdiel, R.~Gopakumar and B.~Knighton,
``Free field world-sheet correlators for ${\rm AdS}_3$,''
JHEP {\bf 2102} (2021) 081 
{\tt [\href{https://arxiv.org/abs/2009.11306}{arXiv:2009.11306 [hep-th]}]}.

\bibitem{Knighton:2020kuh}
B.~Knighton,
``Higher genus correlators for tensionless AdS$_{3}$ strings,''
JHEP \textbf{2104} (2021) 211
{\tt [\href{https://arxiv.org/abs/2012.01445}{arXiv:2012.01445 [hep-th]}]}.

\bibitem{Berkovits:1999im}
N.~Berkovits, C.~Vafa and E.~Witten,
``Conformal field theory of AdS background with Ramond-Ramond flux,''
JHEP {\bf 9903} (1999) 018
{\tt  [\href{https://arxiv.org/abs/hep-th/9902098}{hep-th/9902098}]}.

\bibitem{Gaberdiel:2021iil}
M.R.~Gaberdiel and R.~Gopakumar,
``The String Dual to Free ${\cal N}=4$ Super Yang-Mills,''
{\tt \href{https://arxiv.org/abs/2104.08263}{arXiv:2104.08263 [hep-th]}}.

\bibitem{Gaberdiel:2021jrv}
M.R.~Gaberdiel and R.~Gopakumar,
``The Worldsheet Dual of Free Super Yang-Mills in 4D,''
{\tt \href{https://arxiv.org/abs/2105.10496}{arXiv:2105.10496 [hep-th]}}.

\bibitem{Troost:2011fd}
J.~Troost,
``Massless particles on supergroups and $AdS_3 x S^3$ supergravity,''
JHEP \textbf{1107} (2011) 042
{\tt [\href{https://arxiv.org/abs/1102.0153}{arXiv:1102.0153 [hep-th]}]}.

\bibitem{Gaberdiel:2011vf}
M.R.~Gaberdiel and S.~Gerigk,
``The massless string spectrum on AdS$_3$ x S$^3$ from the supergroup,''
JHEP \textbf{1110} (2011) 045
{\tt [\href{https://arxiv.org/abs/1107.2660}{arXiv:1107.2660 [hep-th]}]}.

\bibitem{Gerigk:2012cq}
S.~Gerigk,
``String States on AdS$_3 \times \rm{S}^3$ from the Supergroup,''
JHEP \textbf{1210} (2012) 084
{\tt [\href{https://arxiv.org/abs/1208.0345}{arXiv:1208.0345 [hep-th]}]}.

\bibitem{gerick:thesis} S.~Gerigk,
``Superstring Theory on $\text{AdS}_3 \times \text{S}^3$ and the $\text{PSL}(2|2)$ WZW Model,''
DISS.\ ETH No.\ 20713 (2012).

\bibitem{Bertle:2020sgd} 
H.~Bertle, A.~Dei and M.R.~Gaberdiel,
``Stress-energy tensor correlators from the world-sheet,''
JHEP {\bf 2021} (2021) 36
{\tt [\href{https://arxiv.org/abs/2012.08486}{arXiv:2012.08486 [hep-th]}]}.

\bibitem{mathematica}
M.~Headrick,
\textit{Virasoro package}.
\href{http://people.brandeis.edu/~headrick/Mathematica/}{http://people.brandeis.edu/~headrick/Mathematica/}

\bibitem{Master}
K.~Naderi, ``Correlators in the Hybrid Formalism'', ETH Master thesis, March 2021.

\bibitem{Dei:2019osr}
A.~Dei, L.~Eberhardt and M.R.~Gaberdiel,
``Three-point functions in AdS$_{3}$/CFT$_{2}$ holography,''
JHEP \textbf{1912} (2019)  012
{\tt [\href{https://arxiv.org/abs/1907.13144}{arXiv:1907.13144 [hep-th]}]}.

\bibitem{Lust:1989tj}
D.~Lust and S.~Theisen,
``Lectures on string theory,''
Lect.\ Notes Phys.\ \textbf{346} (1989).

\bibitem{Dei:2021xgh}
A.~Dei and L.~Eberhardt,
``String correlators on $\text{AdS}_3$: Three-point functions,''
{\tt \href{https://arxiv.org/abs/2105.12130}{arXiv:2105.12130 [hep-th]}}.

\end{thebibliography}
\end{document}